\documentclass[hidelinks,a4paper, 12pt]{article}

\usepackage{latexsym,amsmath,amssymb,graphics,amscd}
\usepackage[colorlinks,
citecolor=blue, linkcolor=black,urlcolor=blue]{hyperref}

\usepackage{hyperref} 
\usepackage[justification=centering]{caption}
\usepackage{stfloats}
\usepackage[longnamesfirst]{natbib}
\usepackage{mathrsfs}
\usepackage{graphicx}
\usepackage{subfigure}
\graphicspath{{Images/}}  
\usepackage{array,tabularx,tabulary,booktabs,arydshln}

\allowdisplaybreaks

\usepackage{stackengine}
\def\Ruble{\stackengine{.67ex}{%
  \stackengine{.48ex}{\textsf{P}}{\rule{.8ex}{.12ex}\kern.6ex}{O}{r}{F}{F}{L}%
  }{\rule{.8ex}{.12ex}\kern.6ex}{O}{r}{F}{F}{L}\kern-.1ex}

\usepackage{array}
\newcolumntype{H}{>{\setbox0=\hbox\bgroup}c<{\egroup}@{}}

\usepackage{longtable}  
\usepackage{multirow} 
\usepackage{lscape}
\usepackage{setspace}

\usepackage{amsmath}
\usepackage{geometry}
\usepackage{soul}

\usepackage{titling}
\usepackage{changes}

\definechangesauthor[name={Alex}, color=red]{A}
\definechangesauthor[name={Mikhail}, color=magenta]{M}

\usepackage{comment}

\usepackage{caption} 
\usepackage[bottom]{footmisc}

\title{
\Large 
\bf Bank Cost Efficiency and Credit Market Structure \\
Under a Volatile Exchange Rate
\thanks{We are grateful to Thorsten Beck (the Editor), the two anonymous referees, Indraneel Chakraborty, Husnu Dalgic, Karligash Glass, Marek Kapicka, Subal Kumbhakar, Michael McCracken, Peter Schmidt, Ctirad Slavik, Martin Uribe, and participants of the 13th International Conference of the ERCIM WG on Computational and Methodological Statistics (CMStatistics 2020, London), 12th (virtual) North-American Productivity Workshop (vNAPW 2021, Miami), 1st Inaugural International Conference on Econometrics and Business Analytics (iCEBA 2021, Saint-Petersburg), and 5th International Conference on Econometrics and Statistics (EcoSta 2022, Japan) for helpful comments and discussion. 
Prokhorov's research was supported by a grant from the Russian Science Foundation (20-18-00365).
}}
\author{\normalsize \textsc{Mikhail Mamonov}\thanks{TBS Business School, Department of Economics and Finance, 1M Place Alphonce Jourdain, 31068 Toulouse France. E-mail: \url{m.mamonov@tbs-education.fr}.} \and \normalsize \textsc{Christopher F. Parmeter}\thanks{Department of Economics, University of Miami, US; e-mail: \url{cparmeter@bus.miami.edu}.} \and \normalsize \textsc{Artem B. Prokhorov}\thanks{Discipline of Business Analytics, University of Sydney, Australia;  e-mail: \url{artem.prokhorov@sydney.edu.au}.
} \thanks{CEBA, St. Petersburg State University, Russia.} \thanks{CIREQ, University of Montreal, Canada.}}

\date{\today}

\begin{document}

\maketitle

\pagenumbering{gobble}

\pagenumbering{arabic}
\setcounter{page}{1}
\vspace{-0mm}
\begin{abstract}
We study the impact of exchange rate volatility on cost efficiency and market structure in a cross-section of banks that have non-trivial exposures to foreign currency (FX) operations. We use unique data on quarterly revaluations of FX assets and liabilities (Revals) that Russian banks were reporting between 2004 Q1 and 2020 Q2. {\it First}, we document that Revals constitute the largest part of the banks' total costs, 26.5\% on average, with considerable variation across banks. {\it Second}, we find that stochastic estimates of cost efficiency are both severely downward biased -- by 30\% on average -- and generally not rank preserving when Revals are ignored, except for the tails, as our nonparametric copulas reveal. To ensure generalizability to other emerging market economies, we suggest a two-stage approach that does not rely on Revals but is able to shrink the downward bias in cost efficiency estimates by two-thirds. {\it Third}, we show that Revals are triggered by the mismatch in the banks' FX operations, which, in turn, is driven by household FX deposits and the instability of Ruble's exchange rate. {\it Fourth}, we find that the failure to account for Revals leads to the erroneous conclusion that the credit market is inefficient, which is driven by the upper quartile of the banks' distribution by total assets. Revals have considerable negative implications for financial stability which can be attenuated by the cross-border diversification of bank assets. 



\end{abstract}
\thispagestyle{empty}
\doublespacing

\newpage
\setcounter{page}{1}
\section{Introduction} \label{sec:Intro}
Substantial variation in banks' engagement in foreign currency operations is evident among emerging market economies (EMEs), and, according to some estimates, foreign currency nominated debt issued by banks' customers in EMEs has quadrupled since 2007 \citep{Acharya21}. 
Though banks attempt to match properly the currency structure of their assets and liabilities \citep{bds2018} and employ forward currency contracts \citep{Ippolito2002} to hedge against foreign exchange (FX) risk, 
currency mismatches have been nonetheless growing across EMEs since 2007 \citep{Kuruc2017,Bruno20}.
\footnote{One of the reasons for growing mismatches is that banks, as is their wont, follow the demand of their customers for foreign-currency deposits, loans, and other services \citep{Brown2011, Luca2008, Ize2003}. And the demand for, say, FX deposits may exceed the demand for FX loans, or vice versa. Another reason is the rising difference between domestic and international prices of borrowed funds for the firms operating in EMEs \citep{Bruno17}.} For banks operating in EMEs, these currency mismatches may create considerable losses during periods of exchange rate instability \cite[see, e.g.,][for evidence from Argentina, Hungary, Russia and Turkey]{Hebert2017, Verner2020, al2019, diGiovanni2022}. 
Though the existing literature vividly shows that currency crises lead to bank runs \citep{Schumacher00} and amplify economic crises \citep{Verner2020}, much less attention has been paid to how exchange rate instability affects the assessment of bank performance and the structure of the credit market in the presence of non-trivial bank exposures to foreign currencies. 

During episodes of exchange rate depreciation, banks bear additional costs due to substantial revaluations of their international and domestic liabilities denominated in foreign currencies. These revaluations become visible in banks' P\&L accounts when banks convert their FX liabilities into domestic currency. These conversions are a part of banks' operating costs, and may thus create a misleading impression that the banks are heavily inefficient, given the same input prices and outputs. 
In this paper, we study how currency revaluations blur estimates of bank cost efficiency and how one can avoid this blurring effect in general. We also explore the drivers of currency revaluations and more general implications of ignoring them. Among the latter, we are specifically interested in credit market efficiency and financial stability when currency revaluations are explicitly taken into consideration.\footnote{We use the term {\it revaluation}, rather than {\it devaluation}, because costs for banks can go up or down, and the negative connotation of devaluation, unlike the more neutral revaluation, misses this point.} 


For our empirical analysis, we appeal to the banking sector of Russia from 2004 to pre-Covid 2020 for the following four reasons. 
First, 
Russia stands as an ``Emerging Market and Developing Economy" with extreme dependence on the export of natural resources \citep{imf:15} and, while the Russian economy grew during 2004--2019 by 3.0\% per annum on average, the banking system's credit to non-financial firms expanded at 8.5\% and to households at 19.5\% (in real terms). 
Second, the Russian economy exhibits high exchange rate volatility, especially after the Russian Ruble was converted to floating in 2013 and the economy encountered a negative oil price shock and Western sanctions in 2014 \citep{al2019} causing a 90\% devaluation of the Ruble. 
Third, the share of FX deposits in Russian banks is as high as 25\% of all deposits and the share of FX loans to firms in non-tradable sectors reaches 50\% of all loans, indicating 
a high level of holdings in foreign currencies on both sides of the balance sheet \citep{bds2018}. 
Fourth, Russian banks have been reporting unique data on the exact amounts of the quarterly revaluations of their FX assets and liabilities on both sides of their P\&L accounts since 2004, thus covering all major episodes of Ruble depreciation before Covid-19. This feature of the data enables testing for the effects of currency revaluations and distinguishes our study from the existing literature on banking in EMEs.


In the first part of our empirical design, we appeal to stochastic frontier analysis and specify a translog cost function with two different versions of the dependent variable. The first is operating costs ($OC$) which, as in previous studies, equals the difference between total costs and interest expenses to eliminate the cross-sectional variation in bank market power when estimating cost efficiency \citep{Berger98}. The second is currency-adjusted operating costs ($CAOC$) which removes FX revaluations ($Revals$, hereinafter) from $OC$. Our estimates indicate that the mean cost efficiency score equals 60\% when using $OC$, and it rises to almost 90\% when employing $CAOC$ (averaged across all banks and quarters of 2005--2020). In both cases, the functional form of the frontier is kept identical. That is, we document a 30 percentage point gap in the estimates of bank cost efficiency stemming from ignoring $Revals$. The blurring effect of $Revals$ is thus huge and may lead to spurious conclusions regarding (comparative) bank performance. 

As a prominent example, we show that one would have to conclude that foreign banks operating in Russia were the least cost efficient group in the banking system, which clearly stands at odds with the literature \citep{LaPorta02, Bonin05, Mian06, Berger09, Beck15}. We show that this spurious conclusion arises because foreign-owned banks, ``by construction," possess the largest portion of FX operations among all banks in the system -- about 36\% of total liabilities (averaged across 2005--2020) -- and therefore $Revals$ crucially affect the assessment of their performance.\footnote{For comparison, domestic banks report only a 10\% share of these operations on average over the same time horizon.} Once we drop $Revals$ from operating costs, we obtain the highest cost efficiency scores for the group of foreign banks in Russia -- even larger than the scores of domestic private banks and leaving far behind the scores of government-owned banks.

Concluding the first part of our analysis, we attempt to generalize our currency-adjusted cost efficiency approach to settings where $Revals$ are not explicitly observed. We perform a counterfactual analysis in which we assume that no such data is available in Russia but we still want to exclude the contaminating effect of its presence. We propose a two-stage modification to the stochastic frontier analysis where the first stage eliminates the macroeconomic effects of exchange rate fluctuations on bank costs. The second stage further accounts for cross-sectional variation in the currency structure of bank assets and liabilities. Strikingly, we show that under our two-stage approach, the mean estimated cost efficiency score reaches 80\%. That is, even without $Revals$, our approach is able to close the 30 percentage point (pp) gap in the cost efficiency scores by almost two-thirds. We believe our approach could apply to banking systems in other EMEs.

In the second part of our empirical analysis, we explore potential channels through which $Revals$ affect the estimated bank cost efficiency scores and the cross-sectional variation in these scores. First, by running a series of two-way fixed-effect regressions, we show that the negative foreign currency mismatch is channeled through households' FX deposits and subsequently translates into larger $Revals$. The relationship is stronger for smaller and less leveraged banks. Second, we show that dropping $Revals$ from $OC$ is not rank preserving: bank rankings in cost efficiency change unpredictably for the majority of banks in the middle of the distribution by the banks' size of total assets (recall the example of foreign banks above). For the left and right tails, we reveal an extreme {\it tail dependence} in the two rankings: applying nonparametric copulas, we find that both extremely inefficient and efficient banks preserve their rankings irrespective of whether $Revals$ are kept or dropped. 

In the third part of our empirical investigation, we analyze broader implications of $Revals$ for the credit market structure. We test the ``efficient structure hypothesis'' (ESH) \cite[see][]{Demsetz73} against the ``quiet life hypothesis'' (QLH) \citep[see, e.g.,][]{Berger98, Koetter12} in the presence of high exchange rate volatility. Our estimates favor ESH in the full sample of banks, meaning that more cost efficient banks are able to earn greater market shares and thus extend more credit to the economy. We show that $Revals$ matter a lot in this direction: in response to a one standard deviation change in the estimated cost efficiency score, corporate credit growth accelerates by 3.7 pp per annum when $Revals$ are dropped and by 2.1 pp when $Revals$ are kept. Using the elasticities of GDP to credit growth from the macro-finance literature \citep{Gambetti2017}, we establish that the 1.6 pp difference in the two estimates is economically significant -- it translates into 0.8 pp of annual GDP growth. 

Furthermore, in the subsample of larger banks, 
keeping $Revals$ inside the estimated cost efficiency scores leads to an erroneous conclusion that the credit market is inefficient, especially for credit to non-financial firms, which is at odds with previous research \citep{Delis12}. Conversely, dropping $Revals$ reverses this finding. Therefore, the outcome of testing for ESH against QLH depends crucially on whether one takes into account $Revals$, or at least the currency structure of bank operations when the exchange rate is highly volatile and banks are sufficiently large. 

Finally, we explore the relationship between $Revals$ and financial stability. Given the negative currency mismatches outlined above, one may anticipate that losses stemming from $Revals$ can create a direct threat to financial stability by reducing bank profitability and also an indirect ``credit risk" threat because monetary authorities may combat the devaluation of local currencies by sustaining higher interest rates in the economy \citep{Goldstein2004}. By employing the Z-score of a bank's distance to default \citep{Beck2013}, we find that a one standard deviation increase in $Revals$ is associated with ({\it i}) a 3.9 point decrease in the Z-score
, and ({\it ii}) a 0.1 log point increase in the (domestic) non-performing loans ratio (NPLs), 
with both effects being economically significant. Exploring cross-sectional variation, we reveal that higher bank capital buffers are not able to smooth the negative association between $Revals$ and bank stability, whereas larger foreign asset holdings are efficient in this direction. 

Our work has several contributions to the empirical banking literature. First, 
we introduce a new perspective on 
bank cost efficiency into the extensive literature on bank performance \citep{Berger98, Feng09, TurkAriss10, Wheelock12, Koetter12, Hughes13, Kumbhakar14a, Spierdijk17, Huang18}. Specifically, we show that one should not confuse bank managers' {\it ability} to conduct operations efficiently with the situation where banks have {\it minor} operations in foreign currencies. A bank is cost efficient if it is able to produce the same amount of outputs with the lowest possible costs, given the factor input prices dictated by the market, the bank managers' risk aversion, {\it and} the currency structure of assets and liabilities under the prevailing exchange rate volatility. Recall that \cite{Hughes13} similarly show that 
greater risk aversion of bank managers does not necessarily imply greater cost inefficiency.

Second, we provide empirical evidence on how the efficiency of financial intermediation evolves across banks and over time when banks operate under a flexible exchange rate regime. As shown by \cite{BenZeev17}, EMEs with flexible regimes face lower output losses when negative credit supply shocks occur. Our findings may provide a channel through which the results of \cite{BenZeev17} work: the lower losses may stem from banks' {\it greater} cost efficiency under flexible regimes so that the banks are {\it better} able to sustain credit to the domestic economy when global credit shocks occur. In the case of Russia, one such global shock was clearly Western sanctions \citep[see, e.g.,][]{al2019}.

Third, our paper shows the importance of currency revaluations for the supply of bank credit in emerging economies. Recently, \cite{Beck2022} show that (German) banks are better able to extend credit to non-financial firms during and after the periods of local currency (euro) depreciation if the banks have positive {\it net} assets in foreign currencies (positive FX mismatch, in our terminology). This is exactly because of additional income -- positive $Revals$ in our case -- that improves the capital adequacy ratio of the banks. In our setting, (Russian) banks operate under negative net assets in foreign currencies and thus their ability to extend credit is limited during the periods of local currency (ruble) depreciation. It is thus important not to overestimate their inability to lend to the economy. Dropping negative $Revals$ from their total costs solves the issue.


\section{Data sources and stylized facts about Russian banks} \label{sec:Data}

\subsection{Exchange rates and bank FX operations across countries}

Table \ref{tab:cross_country_compare} presents summary statistics on FX assets and liabilities of banking systems of the eight selected EMEs and five advanced economies during 2000--2020 against the background of the volatility of their corresponding nominal exchange rates. We observe that banking systems in both EMEs and advanced countries possess similar variations in the magnitude of FX operations (relative to the size of the banking system). However, we  also observe that EMEs encounter {\it far greater} fluctuations in their currency exchange rates. We argue that the precise combination of the two factors is crucial for the assessment of cost efficiency. If, e.g., a banking system holds sufficiently high FX assets and liabilities but the economy's nominal exchange rate is relatively stable, one can ignore the currency structure of bank operations when estimating the cost efficiency. Similarly, if the currency exchange rate is highly volatile but the banking system has few FX holdings, this too will not cause problems for the econometrician evaluating bank cost efficiency. Our point is that it is precisely the combination of a highly volatile exchange rate and non-trivial holdings of FX loans and liabilities that obscures cost efficiency. Our goal is to shed more light on this issue and propose a solution to the problem of estimating bank cost efficiency in countries with exchange rate instability and high dollarization of banking operations. 

\begin{table}[h!] 
\footnotesize
\caption{Banking sector operations in foreign currencies and the volatility of domestic currency exchange rate across countries}\label{tab:cross_country_compare}
\centering
\begin{tabular}{@{} p{0.1\textwidth} @{} 
p{0.20\textwidth} @{} 
>{\centering\arraybackslash}m{0.15\textwidth} @{}
>{\centering\arraybackslash}m{0.15\textwidth} @{}
>{\centering\arraybackslash}m{0.03\textwidth} @{}
>{\centering\arraybackslash}m{0.08\textwidth} @{}
>{\centering\arraybackslash}m{0.08\textwidth} @{}
>{\centering\arraybackslash}m{0.08\textwidth} @{}
>{\centering\arraybackslash}m{0.08\textwidth} @{} }
\toprule[0.5mm]
                       & & \multicolumn{2}{c}{Banking sector: shares of foreign} & & \multicolumn{4}{c}{Nominal exchange rate volatility} \\ 
                       & & \multicolumn{2}{c}{currency operations, \%} & & \multicolumn{4}{c}{GARCH(1,1)}
                       \\
                       \cmidrule(r){3-4} \cmidrule(l){6-9}
                       & & in total loans & in total liabilities & & Mean & SD & Min & Max \\ \hline \noalign{\vskip 2mm} 
\multicolumn{9}{l}{\hspace{-2mm} {\it Panel 1: Selected emerging market economies (EMEs)}:} \\ \noalign{\vskip 2mm}
\hspace{2mm} 1  &  Argentina  &  10.6  &  15.5  &     &  8.7  &  138.2  &  1.2  &  5,313.5 \\
\hspace{2mm} 2  &  Brazil  &  13.8  &  13.3  &     &  9.0  &  31.3  &  0.1  &  813.2 \\
\hspace{2mm} 3  &  Chile  &     &     &     &  2.3  &  4.5  &  0.3  &  112.6 \\
\hspace{2mm} 4  &  China  &     &     &     &  0.1  &  0.4  &  0.0  &  9.6 \\
\hspace{2mm} 5  &  Columbia  &     &     &     &  3.3  &  7.9  &  0.1  &  176.2 \\
\hspace{2mm} 6  &  India  &  8.8  &  8.2  &     &  1.1  &  2.5  &  0.0  &  37.8 \\
\hspace{2mm} 7  &  Russia  &  26.6  &  29.2  &     &  16.0  &  193.7  &  0.0  &  7,854.6 \\
\hspace{2mm} 8  &  South Africa  &  8.2  &  5.5  &     &  5.6  &  11.1  &  0.7  &  251.0 \\
\hspace{2mm} 9  &  Turkey  &  25.9  &  41.4  &     &  13.5  &  122.7  &  0.4  &  3,521.0 \\ 
\hspace{2mm} 10  &  Uruguay  &  63.2  &  70.5  &     &  2.9  &  12.7  &  0.0  &  449.2 \\ \noalign{\vskip 4mm}
\multicolumn{9}{l}{\hspace{-2mm} {\it Panel 2: Selected advanced economies}:} \\ \noalign{\vskip 2mm}
\hspace{2mm} 1  &  Austria  &  21.9  &  13.8  &     &  2.0  &  2.6  &  0.5  &  72.4 \\   
\hspace{2mm} 2  &  Canada  &  26.2  &  40.5  &     &  1.5  &  2.8  &  0.3  &  87.2 \\
\hspace{2mm} 3  &  France  &  8.8  &  16.6  &     &  2.0  &  2.6  &  0.5  &  72.4  \\ 
\hspace{2mm} 4  &  Germany  &  11.0  &  9.3  &     &  2.0  &  2.6  &  0.5  &  72.4 \\
\hspace{2mm} 5  &  United Kingdom  &  55.7  &  53.4  &     &  1.8  &  3.7  &  0.4  &  107.3 \\ \noalign{\vskip 4mm}
 & Time span & \multicolumn{2}{c}{2005--2015} & & \multicolumn{4}{c}{1999M1--2020M11} \\
\bottomrule[0.5mm] \noalign{\vskip 1mm}
 \noalign{\vskip 1mm}
\end{tabular}
\begin{minipage}{1\linewidth}
        {\it Notes}. Nominal exchange rate volatility is estimated using a GARCH(1,1) model \citep{Bollerslev86} on one-week nominal exchange rate returns for each country in the table. \\ {\it Sources}: 1) World Bank global financial database, 2) BIS daily exchange rates statistics (national currency per US dollar).	
\end{minipage}
\end{table}

\subsection{Construction of the panel data on Russian banks}
We gather data on Russian banks from the official database of the CBR. Namely, the data on bank assets and liabilities comes from the monthly balance sheets (Form 101), and the data on revenues and costs are retrieved from the quarterly profit and loss accounts (Form 102), both being publicly disclosed beginning in 2004 and ending in February 2022.
\footnote{See \url{https://www.cbr.ru/banking_sector/otchetnost-kreditnykh-organizaciy/} (in Russian).} The data covers about 95\% of the banking system's total assets. 
With this data at hand, we construct a quarterly panel on bank assets,  liabilities, revenues, and costs over the period 2004 Q1 to 2020 Q2. 

We further merge information on bank ownership status to our quarterly panel: whether a given bank is state-owned or controlled, private domestic or a subsidiary of a foreign bank. The ownership data comes from \cite{Karas2019}. We split the group of state-owned banks into two distinct groups: the so-called ``{\it Big-4}'' and the rest. The Big-4 subgroup comprises the four largest (by assets) banks in the system (Sberbank, VTB, Gazprombank, and Russian Agricultural Bank), which used to be 
parts of the Gosbank of the USSR. 
The rest of the state-owned banks (about 15 credit institutions) are mainly the subsidiaries of the Big-4 or other state corporations. Initially, the group of foreign banks includes about 70 financial institutions, and we then drop those banks that are owned by Russian residents but are headquartered in tax havens outside of Russia. The rest are privately held domestic financial institutions.

At the beginning of the sample period in 2004 Q1, our panel accommodates nearly 1,000 banks. By the end of the sample period in 2020 Q2, the number of banks in the sample shrinks to only 400. \cite{Goncharenko2022} document that this disappearance of banks is a consequence of the bad bank closure policy that had been undertaken by the CBR from mid-2013 to early 2018 and aimed at cleaning the banking system from unfair competitors engaged in balance sheet falsification and capital misreporting. This trend raises an interesting question of how the associated survivorship bias may affect the assessment of bank performance and our estimates of bank cost efficiency. We elaborate on this issue in Section \ref{sec:concerns}.

We clean the data by (i) winsorizing each ratio variable, e.g., equity-to-assets ratio, to be in the interval between the $1^{st}$ and  $99^{th}$ percentile, (ii) winsorizing the annual growth rate of bank total assets to be between the $1^{st}$ and $95^{th}$ percentile (to eliminate M\&A cases), and (iii) dropping all banks operating less than eight consecutive quarters.\footnote{In the sensitivity analysis, we also applied weaker conditions, e.g., requiring banks to operate during at least one quarter, and stricter conditions, e.g., keeping only those banks that survived till the end of the sample period (see Section \ref{sec:concerns}). This did not change the results substantively.} Doing so results in a final sample of \textcolor{black}{1,124} unique banks and a total number of \textcolor{black}{38,484 bank-quarter} observations. 

\subsection{Revaluations of FX assets and liabilities} \label{sec:ConceptOfRevals}
What makes our data appealing is how banks report their financial operations to the CBR. Specifically, in Form 101 banks must disclose the currency structure of their operations in rubles and in foreign currencies, converted to rubles \cite[see, e.g.,][]{Chernykh11}. 
In Form 102 the banks must further reflect the cumulative amount of $Revals$, i.e., the {\it revaluations} generated when they convert their FX-nominated items (loans, eurobonds, etc.) to rubles using the close price of the corresponding currency at the end of the operating day. 
To illustrate the importance of revaluations in the structure of bank costs, we next provide a simple numerical example.

\subsubsection{{\it A numerical example}}
Consider a bank that takes a \$10,000 deposit. At the time of issuance, the $\Ruble$-to-\$ exchange rate is 75-1. At the next instance of reporting, suppose that the amount deposited is still \$10,000, but the $\Ruble$-to-\$ exchange rate is now 100-1 (ruble depreciated). The equivalent value of the deposit reported to the CBR is \Ruble100,000 then, an increase of 33\%. On the costs side of Form 102 the bank would report \Ruble25,000 of additional holdings in the line ``negative revaluations.'' If, conversely, the ruble would appreciate from 75-1 to 50-1, the bank would face a reduction of the deposit equaling \Ruble25,000 and report it in the line ``positive revaluations'' on the revenues side. 
Our example here focused on the liability side. The same holds for the asset side, but in this case, the currency exchange rates are flipped when the ruble appreciates, banks assets are smaller and when the ruble depreciates, assets are larger. So we observe the same data entries but the logic is reversed. Importantly, in all these cases, there are no changes to supply or demand -- only nominal exchange rate fluctuations.


\subsubsection{{\it 
Currency-adjusted operating costs (CAOC)}} 

It is clear from the numerical example above that even two banks of identical size may have substantially different total costs if one has, say, zero operations in foreign currencies and the other has, as the opposite extreme, a 100\% share of such operations. To properly assess the banks' {\it operating} cost efficiency, we need a measure of operating costs, allowing for an apples-to-apples comparison across banks. 

Typically, the literature on bank efficiency suggests subtracting interest expenses ($IE$) from total costs ($TC$), claiming that $IE$ reflects a certain degree of bank market power in the market for borrowed funds rather than managerial abilities to operate the bank \citep{Berger98}. Thus, the resultant operating costs $\overline{OC}=TC-IE$ are used as the dependent variable in analyzing cost efficiency. This procedure works well for economies with stable currencies. However, for EMEs, like Russia, with volatile exchange rates and non-trivial FX operations, $\overline{OC}$ contains, among other things, negative $Revals$, i.e., components that follow exchange rate fluctuations but do not reflect changes in supply and demand.\footnote{$Revals$ are generated on both sides: $Revals^+$ from the income side and $Revals^-$ from the cost side. When the ruble depreciates, FX assets raise $Revals^+$ whereas FX liabilities expand $Revals^-$, holding both FX assets and liabilities fixed (as in the numerical example above).} 
We thus need to remove the costs associated with rising liabilities (e.g., FX deposits) due to ruble depreciation and decreasing assets (e.g., FX loans) due to ruble appreciation:
\begin{align}
        \underline{OC} & = \overline{OC} - Revals^-, \label{eq:ConceptOfRevals}
\end{align}

$\underline{OC}$ is a measure of ``pure'' operating costs independent of the ruble exchange rate fluctuations and which reflects overhead costs, i.e., the costs associated with assessing borrowers' quality, the costs of attracting deposits, etc. To avoid confusion, from now on we refer to $\underline{OC}$ as currency-adjusted operating costs $CAOC$.

\subsubsection{{\it Summary statistics and stylized facts}}
To illustrate the importance of $Revals$ and the computation of $CAOC$ using Eq.~(\ref{eq:ConceptOfRevals}), Table \ref{tab:desc_stats_costs} reports absolute magnitudes of costs and their components across banks averaged across the sample period. The mean $TC$ equals \Ruble94.5 bn (\$3.6 bn at the average exchange rate), 
interest expenses on domestic and foreign liabilities absorb only \Ruble3 bn, and $Revals^-$ account for \Ruble75 bn ranging from nearly \Ruble0 to more than \Ruble120 tr (equivalent to 200\% of Russia's average GDP for the period).\footnote{This extreme case pertains to Sberbank, Russia's largest bank.} Therefore, the operating costs $\overline{OC}$ equal \Ruble91.7 bn, but once $Revals^-$ are further subtracted, the resultant $CAOC$ shrinks to just \Ruble17.3 bn. This leads to the first stylized fact:

\bigskip 
{\bf Stylized Fact \#1}: 
{\it Negative revaluations of FX assets and liabilities constitute the largest portion of banks' total costs in Russia, dramatically exceeding banks' interest expenses.} \vspace{2mm} 

\doublespacing

\begin{table}[h!] 
\caption{Selected components of income and costs: descriptive statistics \\ for the  2005 Q1 -- 2020 Q2 period, billion of rubles ($N=38,484$) }\label{tab:desc_stats_costs}
\centering
\footnotesize
\begin{tabular}{lHcccc}
\toprule[0.5mm]
	                   & Obs & Mean & SD & Min & Max \\ 
	                  \noalign{\vskip 2mm} \hline \noalign{\vskip 2mm} 
\noalign{\vskip 2mm}
(1) \hspace{0mm} Total costs (TC)  &	38,656 & 94.5 &		2,105.7 &	0.001	& 132,126.9 \\ \noalign{\vskip 2mm}
(2) \hspace{4mm}    Interest income (II)     & 38,655  &  5.4 & 58.6 & 0.0001 &  2,203.7 \\
(3) \hspace{4mm}    Interest expenses (IE)     &	38,656 & 2.8  & 	27.1    &	0.00002 &	1,131.3 \\
(4) \hspace{4mm}    $Revals^+$ &	38,656 & 74.5 &		1,889.4	&     0 &		120,119.8 \\
(5) \hspace{4mm}    $Revals^-$ &	38,656 & 74.6 & 1,891.9 &	0 &	120,807.9 \\ \noalign{\vskip 2mm}
(6) \hspace{0mm} Operating costs (OC): (1)-(3) & 38,656  &  91.7 & 2,082.5 & 0.001  & 131,318.8	\\ 
(7) \hspace{0mm} Currency adjusted operating costs (CAOC): (1)-(3)-(5) &	38,656 &  17.3  &  207.1  &  0.001  &  11,199.0 \\ \noalign{\vskip 1mm}
\bottomrule[0.5mm]
\end{tabular}
\end{table}

\begin{figure}[h!]
    \centering
    \subfigure[$Revals^-$, as \% of total costs]{
    \includegraphics[width=0.47\textwidth]{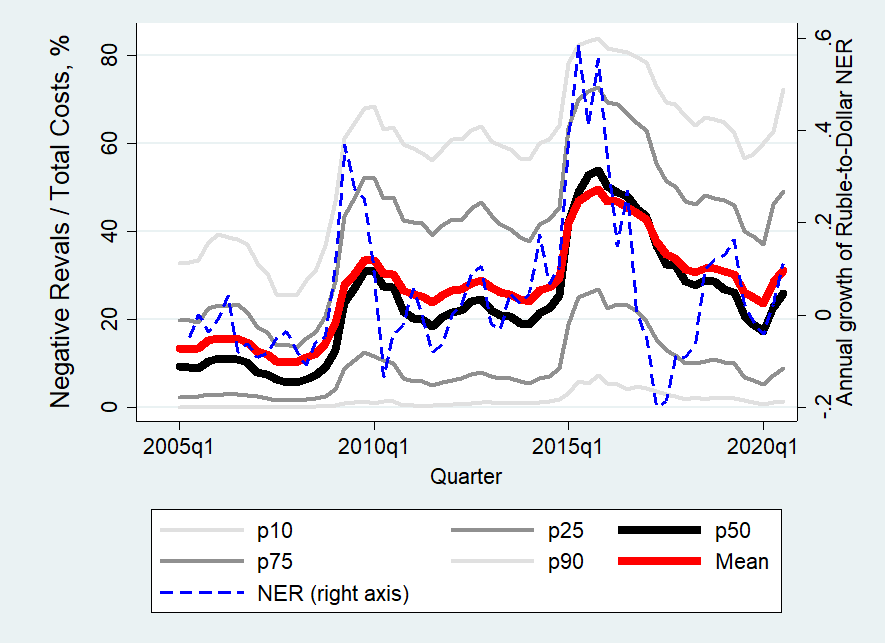}}
    \centering
    \subfigure[$\Delta Revals$, as \% of total costs]{
    \includegraphics[width=0.47\textwidth]{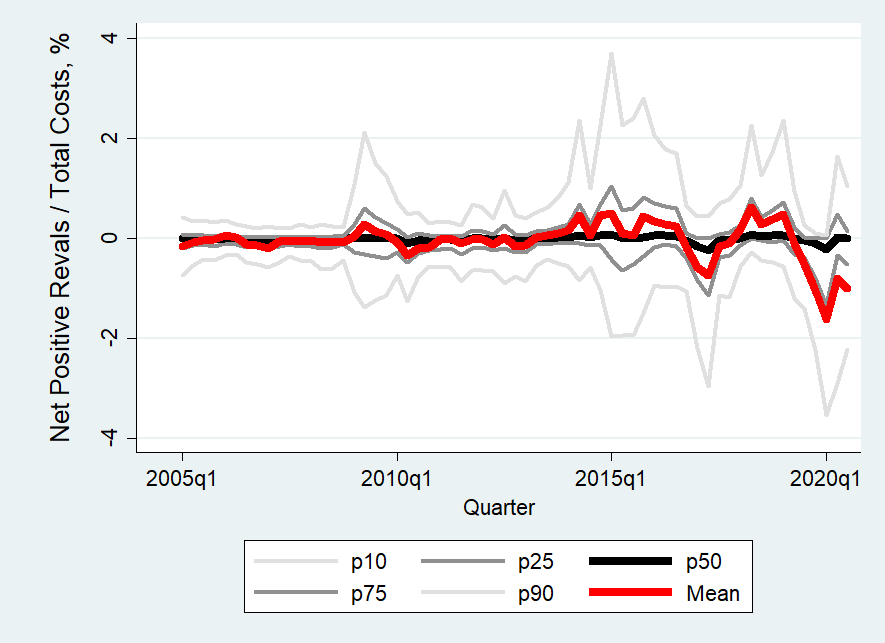}}

    \captionsetup{justification=centering,margin=2cm}
    \caption{Time evolution of banks' positive and negative revaluations of assets and liabilities denominated in foreign currencies}
    \label{fig:Revals}
\end{figure}

Figure \ref{fig:Revals}({\it a}) plots the time evolution of the ratio of $Revals^-$ to TC (at the mean, median, and upper/lower quartiles/deciles) against the annual dollar to ruble nominal exchange rate (NER). The average share of $Revals^-$ in total costs equals 26\% across all banks and quarters, with a rising trend over time and two distinct spikes -- by 20 pp at the start of the Global Financial Crisis of 2007--2009 and by 30 pp in 2014--2015 when both global sanctions and the collapse of the world oil prices hit Russia. Similar patterns are observed in all percentiles of the banks' distribution by the $Revals^-$ to TC ratio. This brings us to the second stylized fact: 

\bigskip
{\bf Stylized Fact \#2}: 
{\it Negative revaluations of FX assets and liabilities are strongly correlated with ruble depreciation against the US dollar during periods of negative external shocks.} \vspace{2mm} 

\doublespacing

Figure \ref{fig:Revals}({\it b}) presents the ratio of $\Delta Revals = Revals^+ - Revals^-$ to $TC$ across the sample period (at the mean, median and upper/lower quartiles/deciles). The mean $\Delta Revals$ ratio was near zero before 2014, meaning that, on average, banks were {\it naturally} hedged against the FX risk in Russia.\footnote{Recall also from Table \ref{tab:desc_stats_costs} that $\Delta Revals$ is just \Ruble0.1 bn on average and compare it to the net interest income $NII=II-IE$ which equals \Ruble2.6 bn and is thus much larger, despite each of the two components in $NII$ being dramatically lower than either of the two components in $Revals$.} However, $\Delta Revals$ largely deviates from zero when the Ruble depreciates against major world currencies: in 2009, in response to the Global Financial Crisis; in 2014--2015, when oil prices collapsed and Russia encountered Western sanctions, and during the first half of 2020 due to the onset of the Covid-19 pandemic. This brings us to the third stylized fact:  

\bigskip

{\bf Stylized Fact \#3}: {\it On average, the banking system is naturally hedged against the FX risk ($\Delta Revals \sim 0$), with considerable cross-sectional variation during periods of negative macroeconomic shocks.} 
\vspace{2mm} 

\doublespacing

It is precisely these higher order moments of $\Delta Revals$ to total costs in Stylized Fact \#3 that are important for the analysis in our paper. Without large variability in $\Delta Revals$ there would be little empirical identification power in quarter-to-quarter changes available. And it is this variation that we exploit, especially in times of negative macroeconomic shocks, to demonstrate the impact that \textit{Revals} have in the cross-section of Russian banks.

\begin{figure}[h!]
    \centering
    \subfigure[FX operations, as \% of total assets]{
    \includegraphics[width=0.47\textwidth]{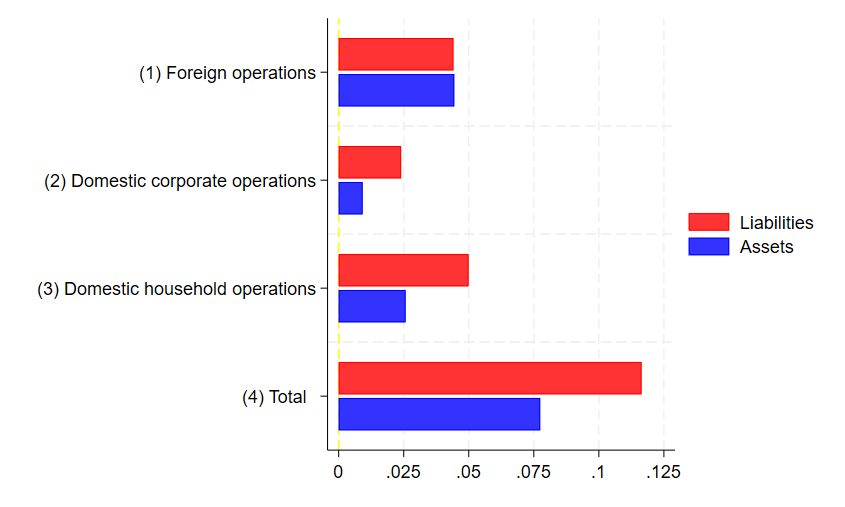}}
    \centering
    \subfigure[Growth rates of FX operations, \% YoY]{
    \includegraphics[width=0.47\textwidth]{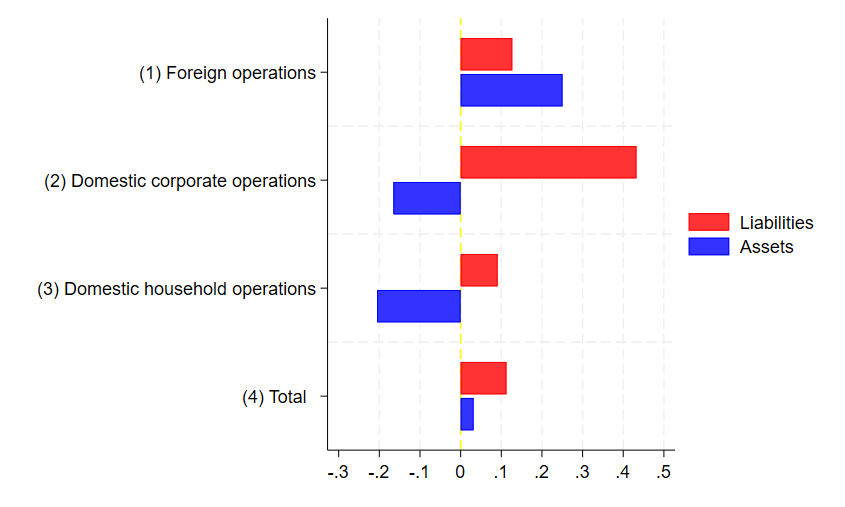}}

    \captionsetup{justification=centering,margin=2cm}
    \caption{Structure and dynamics of FX operations by Russian banks}
    \label{fig:desc_stats_FX}
\end{figure}

We now consider the structure and dynamics of FX operations of Russian banks to understand which types of {\it within-} and {\it outside-}country operations are responsible for generating such large amounts of $Revals^+$ and $Revals^-$ on banks' P\&L accounts. Figure \ref{fig:desc_stats_FX}({\it a}) shows that on average 12\% of banks' total liabilities and 7.5\% of banks' total assets are denominated in foreign currencies, with roughly half of FX liabilities originating from abroad ({\it foreign liabilities}) and two-thirds of FX assets located abroad ({\it foreign assets}). 
While foreign assets and foreign liabilities are perfectly matched, domestic FX operations are, by contrast, largely mismatched, with household FX deposits being two times greater than FX loans to the same category of customers; 
the same is true for firms.
\footnote{During the sample period, Russia's banks had been facing a growing supply of FX funds by local private customers who were lacking trust in local currency and anticipated its further depreciation. Indeed, over 2007--2020, ruble depreciated more than twofold -- from nearly 18 to 49 rubles per US dollar -- and the USD amount of private FX deposits soared from \$26 to \$126 billion.}

Turning to the growth rates in Figure \ref{fig:desc_stats_FX}({\it b}), we observe that Russian banks expand substantially their FX operations abroad, both foreign liabilities and foreign assets (by 12\% and 24\%, respectively), whereas domestically they encounter huge reduction of FX lending to both firms (--18\%) and households (--20\%) on the background of a dramatic rise of FX corporate deposits (42\%) and a moderate rise of FX private deposits (10\%). 

The numbers presented in Figure \ref{fig:desc_stats_FX} imply that an average bank in Russia encounters a {\it negative} FX mismatch: its liabilities denominated in foreign currencies exceed its FX assets. Further computations reveal that the negative FX mismatch holds for roughly 90\% of all banks in Russia. This is largely in line with the situation in other EMEs \citep{Lane2010,Benetrix2015}.\footnote{An interpretation may be that banks largely engage in the cross-border diversification of their operations, and this engagement is likely to be a natural response to the highly volatile macroeconomic and political environment in Russia \citep{Mironov2013,Guriev2019}. Since the first episode of sanctions in 2014, banks have been steadily encountering a rising demand for FX deposits from local customers \citep{Mamonov2023}, which likely reflects a loss of trust in the ruble. Finally, the de-dollarization of domestic credit is a natural demand-side factor that arose in response to the Global Financial Crisis of 2008--2009 in Russia, as it was the case in, e.g., Hungary \citep{Verner2020}, when a large number of local customers with earnings in local currency were unable to repay their FX credits.} 
We can thus formulate two additional stylized facts:  \vspace{4mm} 

\onehalfspacing {\bf Stylized Fact \#4}: {$Revals^-$ are correlated with the negative currency mismatch that Russian banks are facing over time (FX assets are less than FX liabilities).} \vspace{2mm} 

\vspace{2mm}

\onehalfspacing {\bf Stylized Fact \#5}: {\it The negative currency mismatch is driven by domestic FX deposits of households and, to a lesser extent, foreign operations.}\footnote{To provide further evidence of this point, we ran a two-way fixed effects regression of $Revals^-$ on the four types of domestic and two types of foreign FX operations, controlling for bank size, capitalization, and liquidity. Online Appendix \textcolor{red}{A} contains the results.}  

\doublespacing

Other bank-level characteristics relevant to stochastic cost efficiency analysis are summarized in Table \ref{tab:desc_stats} in \ref{app:descstats}. Online Appendix \textcolor{red}{B} contains a comparative analysis of the summary statistics on Russian banks and their corresponding analogs for commercial banks in some other EMEs and the US. 

\section{Estimation of bank cost efficiency} \label{sec:CostEfficiency}
\subsection{Methodology}
Following the vast literature on bank cost efficiency  \cite{Bonin05,TurkAriss10,Tabak12,Hughes13}, we employ stochastic frontier analysis in the spirit of \cite{ALS:77} and we extend the concept of risk-adjusted bank efficiency put forward in \cite{Hughes96}. Specifically, we explicitly introduce the currency structure of bank assets and liabilities (idiosyncratic component) and exchange rate fluctuations (aggregate component) which jointly give rise to heterogeneous $Revals$ at the bank level. We thus define cost efficiency as a bank's ability to operate under lower costs net of $Revals$ given (i) the volume of outputs it produces ($Y_{it}$), (ii) factor input prices it faces in the market ($w_{it}$) and (iii) the risk aversion of bank managers.
The resultant translog cost function is:
\begin{align} \label{eq:RusBanksCostFrontier}
    \ln COST_{it} = \alpha_i &+ f(Y_{it}, w_{it}, Risk_{it}, t) + u_{it} + v_{it}, 
\end{align}
where for bank $i$ in quarter $t$ we consider two versions of the dependent variable $\ln COST_{it}$: one based on $CAOC$ and the other on $\overline{OC}$, as in previous studies. $f(\cdot)$ is the translog cost frontier, which can be expressed in terms of the four arguments, their squared terms, and all their interactions with each other. In the composite regression error term, $u_{it}$ is the cost inefficiency component which is non-negative and assumed to be distributed as Truncated Normal, and $v_{it}$ is an idiosyncratic error.

Following the production approach, we include three outputs ($L=3$): loans ($Y_1$), borrowed funds ($Y_2$), and off-balance sheet activities ($Y_3$).\footnote{The same approach is applied in \cite{Tabak12} to study the efficiency of financial intermediation in Latin America's countries. Note that \cite{Hughes13} consider loans as output and deposits as {\it input factor utilization} while including the price of deposits in the same translog cost function. We also followed that line and obtained the same results as in our baseline.} Loans accommodate domestic credit to non-financial firms, households, and other banks and international credit outside Russia.\footnote{We also estimated a model in which we decomposed loans into more specific categories. Our baseline results remain qualitatively unchanged.} The same structure applies to borrowed funds which consist of funds from domestic non-financial firms, households, and other banks and international borrowings. We include two factor input prices ($M=2$): the price of labor (the ratio of personnel expenses to total assets, $w_1$) and the price of physical capital (expenses unrelated to interest and personnel over total assets, $w_2$). To ensure constant return to scale in prices, we impose standard restrictions on the corresponding coefficients in our translog cost function. 

The banking literature traditionally considers bank {\it ownership status} as a key driver of (in)efficiency concluding that foreign-owned banks outperform domestic banks, both state- and privately-held \citep{Bonin05}. 
To allow for shifts in average inefficiency due to differences in ownership and other characteristics, we follow \cite{Battese95} and assume that $u_{it}\sim N_+\left(\mu(z_{it};\delta),\sigma^2\right)$ where  
\begin{align} \label{eq:RusBanksCovariates}
    \mu(z_{it};\delta) = & \delta_0 + \delta_1 Big\text{-}4_{i} + \delta_2 Other\,State_{it} + \delta_3 Foreign_{it} + \sum^{K}_{k=1} \delta_{3+k} X_{k,it},  
\end{align}
where $Big\text{-}4_i$, $Other\,State_{it}$, and $Foreign_{it}$ are indicator variables equal to one if bank $i$ is a {\it Big-4} state-owned bank, another state-owned or -controlled bank, or a foreign subsidiary, respectively (domestic privately-held banks are set as the reference group). $\mathbf{X}_{it}$ includes the bank liquidity ratio, the ratio of long-term loans to non-financial firms and to households, the annual growth of total assets, equity capital to total assets ratio, and two indicator variables reflecting (i) whether a bank is under a bail-out scheme by the CBR and (ii) whether $t \geq $ Q3 2013, 
i.e., the period of the bad bank closure policy \citep{Goncharenko2022}. 

Given the observed differences in the currency compositions of balance sheets across banks in Russia, it is important to note that the regulatory requirements on ({\it i}) the fraction of funds that must be reserved when extending credit in foreign currencies and on ({\it ii}) the maximum permitted size of the open position in foreign currencies are {\it uniform} across all bank ownership types in Russia, whether public or private, foreign or domestic-owned.\footnote{Reserve requirements on FX operations are summarized here:  \url{https://www.reuters.com/article/orubs-cbr-reserves-changes-idRUKCN0ZD1UG}. According to the Central Bank of Russia's (CBR) ``Instruction 178-I'' dated 28 December 2016, a bank's open FX position cannot exceed 20\% of the bank's total assets.} This simplifies the comparison of bank efficiency across ownership types \citep{Bonin05}.

To avoid estimation bias caused by the non-zero mean of $u_{it}$ \citep{wang_schmidt:2002}, we simultaneously estimate Equations (\ref{eq:RusBanksCostFrontier}) and (\ref{eq:RusBanksCovariates}).  
After estimation, we compute efficiency scores for each bank $i$ in every quarter $t$ as $CE^{(\ell)}_{it} = E\left[e^{-u^{\ell}_{it}}|\epsilon= e_{it}\right]$, 
where 
$\ell=1$ when $Revals^-$ are dropped and $\ell=2$ when they are kept. 

Mechanically (because $f(\cdot)$ is the same for $\ell=1$ and $\ell=2$), 
$CE^{(1)}_{it} \geq CE^{(2)}_{it}$, with strict equality if a bank has no operations in foreign currency. 
However, 
it is not obvious ex-ante whether the exclusion of $Revals^-$ from operating costs is rank preserving: i.e., whether the {\it rankings} of banks in $CE^{(1)}_{it}$ is the same as in $CE^{(2)}_{it}$. 
To investigate this issue, we appeal to copula functions which are a popular tool to capture rank dependence across random variables \cite[see, e.g.,][]{nelson:06}.\footnote{We provide details on copulas in 
Appendix \textcolor{red}{C} of the Supplementary Material.} 

\subsection{Baseline estimation results} \label{sec:CostEfficiency_Results}
\subsubsection{{\it Cost frontier and mean (in)efficiency}} \label{sec:CostEfficiency_Results_Regression}
Table \ref{tab:RusBanksCostFrontier_MeanIneffEquation} presents estimates of the cost frontier and mean inefficiency equations in Panels 1 and 2, respectively. In columns (1) and (2), 
$Revals^-$ are kept inside the operating costs 
and in columns (3) and (4), they are dropped. Columns (1) and (3) contain only bank ownership statuses as explanatory variables in the mean inefficiency equation. Columns (2) and (4) add the rest of the control variables discussed above.

\begin{table}[h!] 
\footnotesize
\caption{Joint estimation results of cost frontier and inefficiency equations}\label{tab:RusBanksCostFrontier_MeanIneffEquation}
\centering
\begin{tabular}{@{} p{0.45\textwidth} @{} 
>{\centering\arraybackslash}m{0.12\textwidth} @{} 
>{\centering\arraybackslash}m{0.12\textwidth} @{} 
>{\centering\arraybackslash}m{0.05\textwidth} @{} 
>{\centering\arraybackslash}m{0.12\textwidth} @{} 
>{\centering\arraybackslash}m{0.12\textwidth} @{}}
	\toprule[0.5mm]

\hspace{45mm} $Revals^-$:      & \multicolumn{2}{c}{Kept}   
            & & \multicolumn{2}{c}{Dropped} 
	          \\ \cmidrule(r){2-3} \cmidrule(l){5-6}
            & Ownership & Ownership + Controls & & Ownership & Ownership + Controls ({\it baseline})
	          \\ \cmidrule(r){2-6}
            &  (1) & (2) & & (3) & (4)  \\ \noalign{\vskip 1mm} \hline \noalign{\vskip 4mm}  

\hspace{0mm}\textit{Panel 1: Cost frontier equation} &  &  &&  &  \\ \noalign{\vskip 3mm}
\hspace{2mm} Constant return to scale & Yes & Yes && Yes & Yes \\ \noalign{\vskip 3mm}
\hspace{2mm} Cost frontier covariates &  Yes & Yes && Yes & Yes \\ \noalign{\vskip 3mm}
\hspace{2mm} SD of inefficiency term ($\sigma_u$) & 1.860*** & 0.939*** && 1.097*** & 1.368*** \\ 
                                                  & (0.087)  & (0.014)  && (0.020)  & (0.108) \\ \noalign{\vskip 1mm}
\hspace{2mm} SD of i.i.d. term ($\sigma_v$) & 0.085*** & 0.061*** && 0.040*** & 0.037*** \\ 
                                            & (0.002)  & (0.001)  && (0.000)  & (0.000)  \\ \noalign{\vskip 6mm}
                                          
\hspace{0mm}\textit{Panel 2: Inefficiency equation} &  &  &&  &                                     \\   \noalign{\vskip 3mm}
\hspace{2mm} Group 1: Domestic private banks & {\it Reference} & {\it Reference} && {\it Reference} & {\it Reference} \\ \noalign{\vskip 3mm}

\hspace{2mm} Group 2: Big-4 banks & 5.122*** & 0.971***&& 1.579**   & 1.094 \\ 
                                  & (0.525)  & (0.098) && (0.697) & (0.934)  \\ \noalign{\vskip 1mm}
\hspace{2mm} Group 3: Other state-owned banks & 3.200*** & 0.715*** && 3.531*** & 3.594*** \\                                                  & (0.329)  & (0.066)  && (0.339)  & (0.706)  \\ \noalign{\vskip 1mm}
\hspace{2mm} Group 4: Foreign-owned banks & 6.027*** & 1.519*** && --2.437*** & --4.789*** \\ 
                                          & (0.499)  & (0.043)  && (0.307) & (0.825)  \\ \noalign{\vskip 3mm}
                        
\hspace{2mm} Bail-out    & & 0.077 && & 2.839*** \\ 
                         & & (0.088)  && & (0.720)  \\ \noalign{\vskip 1mm}
\hspace{2mm} Tight prudential regulation    &  & 1.165*** &&  & 2.196*** \\ 
                                            &  & (0.036)  &&  & (0.381)  \\ \noalign{\vskip 1mm}
\hspace{2mm} Liquidity ratio    &  & --0.007*** &&  & --0.112*** \\ 
                                &  & (0.001)    &&  & (0.020) \\ \noalign{\vskip 1mm}
\hspace{2mm} Long-term loans to firms ratio    &  & 0.013***   && & --0.007 \\ 
                                               &   & (0.002) &&  & (0.013) \\ \noalign{\vskip 1mm}
\hspace{2mm} Long-term loans to households ratio &  & --0.038*** &&  & --0.113*** \\ 
                                                 &   & (0.002)   &&  & (0.020) \\  \noalign{\vskip 1mm}
\hspace{2mm} Annual growth rate of total assets  &  & --0.008*** &&  & --0.027*** \\ 
                                                 &  & (0.000)    &&  & (0.005) \\ \noalign{\vskip 1mm}
\hspace{2mm} Equity capital to total assets ratio    & & --0.018*** &&  & 0.184*** \\ 
                                                     & & (0.001)  &&  & (0.029) \\ \noalign{\vskip 1mm}
\hspace{2mm} Constant    & --7.469*** & --0.475*** && --10.391*** & --13.750*** \\ 
                         & (0.786)   & (0.060)    && (0.380)  & (2.222)  \\ \noalign{\vskip 3mm}

Obs   & 38,484 & 38,484 && 38,484 & 38,484 \\
Banks & 1,124  & 1,124  && 1,124  & 1,124  \\
log likelihood & --13,751 & --10,920 && 34,334 & 35,300 \\
Convergence achieved & Yes & Yes && Yes & Yes \\
\bottomrule[0.5mm] \noalign{\vskip 2mm}
\end{tabular} 
	\begin{minipage}{1\linewidth}
    {\it Note}: ***, **, * indicate that a coefficient is significant at the 1\%, 5\%, 10\% level, respectively. Standard errors are clustered at the bank level and appear in the brackets under the estimated coefficients.   \\		
	\end{minipage}
\end{table}

{\it Cost frontier results}. Describing the estimated coefficients on each and every variable and their products in the translog cost equation goes beyond the scope of our study. We, however, briefly analyze the estimated return to scale ($RTS^{\ell}_{it}$, hereafter) as it provides an aggregate characteristic of the banks' potential to expand outputs given the same costs and risk aversion.\footnote{We computed $RTS_{it}$ as an inverse of the sum of the three estimated output elasticities. We report the resultant time evolution of RTS in Online Appendix \textcolor{red}{D}, see Fig. \textcolor{red}{D.I}.} First, the magnitude of estimated $RTS^{\ell}_{it}$ ranges between 1.2 and 1.3 which is in line with a variety of banking studies across different countries \citep{Feng09, Wheelock12, Hughes13, Kumbhakar14b}. Second, the estimated $RTS^1_{it}$ appears similar to $RTS^2_{it}$. The similarity of estimated $RTS_{it}$ regardless of $Revals^-$ indicates that $Revals^-$ has no impact on the {\it shape} of the cost frontier,\footnote{Pure operating costs $\underline{OC}$ include pay to personnel and capital expenses. These costs do not depend on $Revals^-$ (recall Stylized fact \#2 that $Revals^-$ are highly correlated with the Ruble's nominal exchange rate). Hence it is not surprising that estimated $RTS_{it}$ is consistent between the two competing measures of costs.} but they still may affect the distance to it.

{\it Mean inefficiency results}. Our estimates in columns (1) and (2) show that the foreign-owned banks have the largest positive coefficients across all banking groups in Russia (significant at 1\%). This suggests that keeping $Revals^-$ clearly leads to the controversial conclusion that foreign-owned banks are the least cost efficient financial firms in the banking sector. 
Strikingly, as our estimates in columns (3) and (4) indicate, dropping $Revals^-$ reverts the results: foreign-owned banks turn out to be the only banking group for which we obtain a negative and significant estimate (at the 1\% level), meaning they are more cost-efficient than the reference group, i.e., domestic private banks.\footnote{In Online Appendix \textcolor{red}{E}, we document that foreign-owned banks, together with the Big-4 state-owned banks, have the largest share of $Revals^-$ in their total costs among all bank groups (Figure \textcolor{red}{E.I}). In this situation, it is clear why dropping $Revals^-$ so dramatically affects the conclusion regarding foreign-owned banks. 
} And this conclusion is exactly in line with the legacy of previous studies on the ownership--efficiency nexus \citep{LaPorta02, Bonin05, Berger09, Beck15, Delis16}.\footnote{The existing research has vividly shown that foreign-owned banks, by delivering technological enhancements into domestic banking systems, are likely to be the most cost efficient group. In addition, as \cite{Delis16} reveal, foreign banks worldwide decrease marginal costs in the host banking systems and, through that, can earn greater market power.} 

Regarding the other banking groups, we obtain an insignificant coefficient estimate for state-owned banks and a positive and significant (at 1\%) estimate for the other (smaller) state banks in Russia. This result implies that Russia's {\it Big-4} state-owned banks are as cost efficient as privately-held financial institutions.\footnote{Though it is not our direct focus here, this result may be rationalized through the following facts. First, over the seventy years of the Soviet Union, these four banks (being united in a single institution, the Gosbank of USSR) were the only financial entities operating in the economy. As \cite{Bircan2020} reveal, the regions in Russia with greater intensities of these networks enjoy higher innovative activities of local non-financial firms. Second, given their recognizable brands and explicit support from the government, the Big-4 banks enjoy greater trust from their customers as the customers perceive them as safe havens during (frequent) periods of macroeconomic instability in Russia.} 

Overall, our results indicate that $Revals^-$ are a crucial component of total costs which may contaminate the qualitative outcome of the cost efficiency analysis. We suspect the same can be true for other EMEs with sufficiently volatile exchange rates (recall Table \ref{tab:cross_country_compare}).\footnote{in Online Appendix \textcolor{red}{F}, 
Table \textcolor{red}{F.1} clearly shows that the result on the superiority of foreign-owned banks requires not only dropping $Revals^-$ but also accounting for managerial risk aversion. Conversely, as we show in column (2) of Table \textcolor{red}{F.1}, if we keep interest expenses in the total cost and thus contaminate the estimates of cost efficiency by market power effects \citep{Berger98}, the conclusion that foreign-owned banks are the most cost efficient group would remain unaltered.} 


\subsubsection{{\it Time evolution of the estimated cost efficiency}} \label{sec:scores}

\begin{figure}[h!]
    \centering
    \subfigure[$Revals^-$ kept]{
    \includegraphics[width=0.47\textwidth]{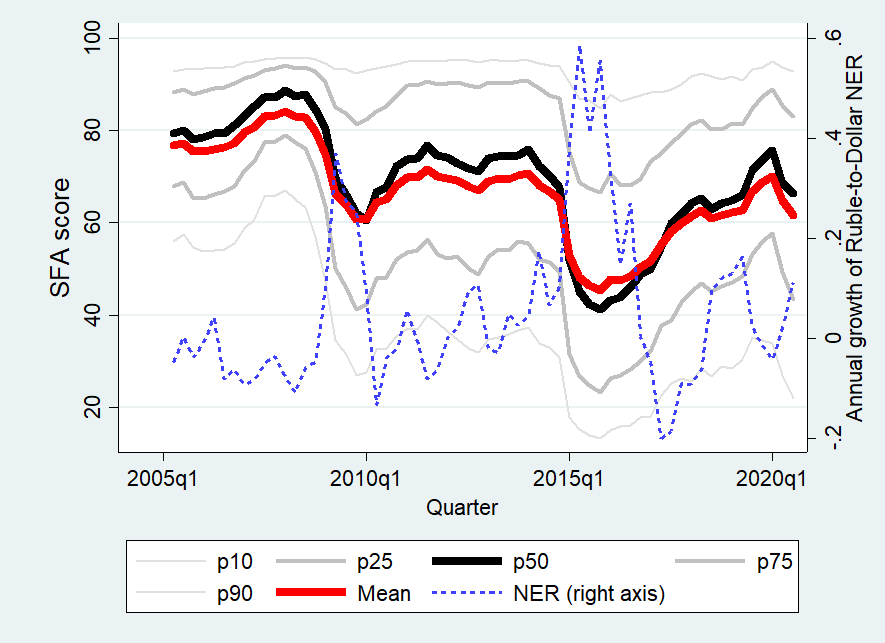}} 
    \subfigure[$Revals^-$ dropped]{
    \includegraphics[width=0.47\textwidth]{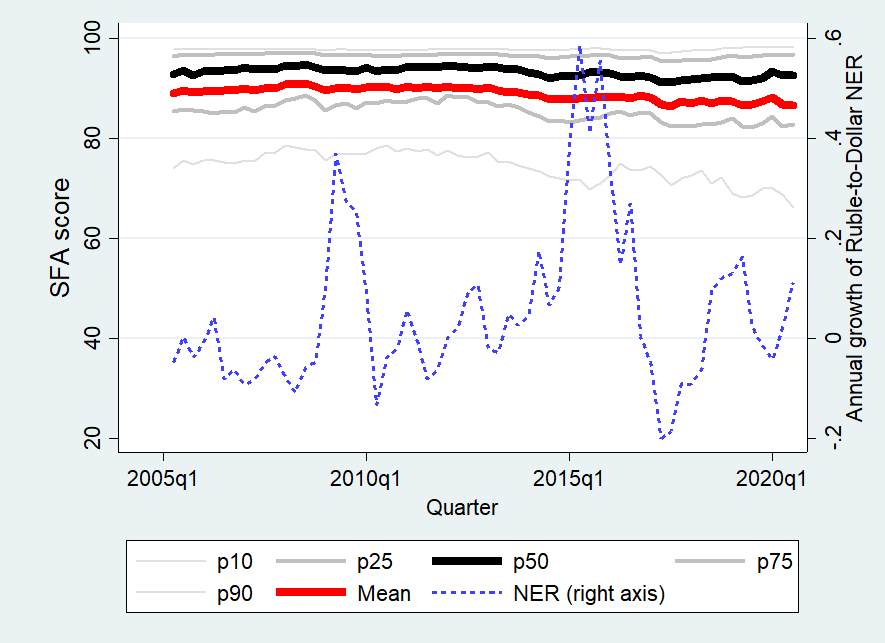}}
    

    \captionsetup{justification=centering,margin=2cm}
    \caption{Time evolution of the estimated cost efficiency (cost efficiency scores): $Revals^-$ kept (a) or dropped (b)}
    \label{fig:SFA_TimeEvol}
\end{figure}

Figure \ref{fig:SFA_TimeEvol} presents the monthly dynamics of the estimated cost efficiency scores (at mean, median, upper/lower quartiles/deciles) when $Revals^-$ are kept ({\it a}) or dropped ({\it b}) on the background of the Ruble-to-US Dollar nominal exchange rate. We observe several broad patterns in the time evolution of the estimated efficiency scores. First, when $Revals^-$ are kept ({\it a}), cost efficiency scores are much more volatile across banks and in time compared to the case when $Revals^-$ are dropped ({\it b}). Moreover, the time pattern itself is reminiscent of the share of $Revals^-$ in total costs (recall Fig. \ref{fig:Revals}({\it a})) which, in turn, is highly correlated with the Ruble's exchange rate (recall Stylized Fact \#2). That is, keeping $Revals^-$ produces a strong cyclical component in the time evolution of cost efficiency scores. This cyclical component, if taken seriously, would imply that the managerial ability to run a bank -- the stock of bank managers' human capital, to push it to an extreme -- shrinks dramatically during times of (currency) crises. Clearly, this contradicts economic intuition. 

Second, the estimate cost efficiency scores turn out to be remarkably stable at the mean, median, and upper quartile when $Revals^-$ are dropped ({\it b}), with only a slight uptick in volatility for the mean and median after 2014. The cyclical component -- and the correlation with the Ruble's nominal exchange rate -- fully disappears for the upper 50\% of observations (i.e., for the more efficient banks). 

Third, dropping $Revals^-$ from operating costs raises the estimated efficiency scores by nearly a half -- from 60\% to 90\%. This 30 pp gap is economically large and important and fully explained by the $Revals^-$ component of bank total costs. Overall, $Revals^-$ may dramatically contaminate the estimates of bank cost efficiency.

\subsubsection{\it Addressing concerns on the survivorship bias} \label{sec:concerns}

\begin{figure}[h!]
    \centering
    \subfigure[$Revals^- / TC$]{
    \includegraphics[width=0.47\textwidth]{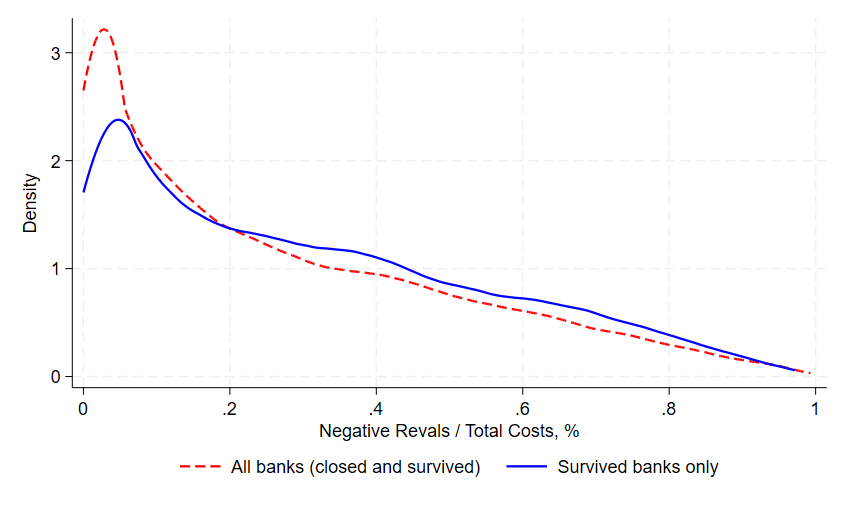}} 
    \subfigure[Net FX position]{
    \includegraphics[width=0.47\textwidth]{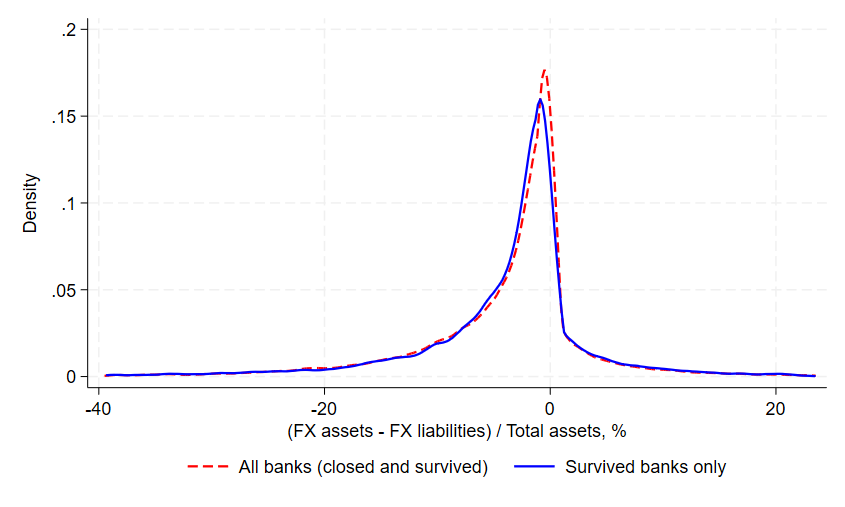}}
    
    \captionsetup{justification=centering,margin=2cm}
    \caption{Kernel densities of $Revals^-$ and net FX position in the full sample and across survived banks}
    \label{fig:kdensity}
\end{figure}

An important concern about our baseline results is a potential survivorship bias due to the bad bank closure policy implemented by the Central Bank of Russia in 2013--2018 which resulted in roughly two-thirds of all operating banks being preemptively closed for financial fraud \citep{Goncharenko2022}. We address this concern by running additional regressions in which we force the sample to contain only those 236 banks that survived the policy and were active during the entire period of 2005 Q1 to 2020 Q2. In this case, the number of bank-quarter observations drops by a factor of 3 compared to the baseline (from 38,484 to 12,950). We re-estimate the translog cost function and present the resulting time evolution of the estimated ``survivorship-robust" cost efficiency scores in Online Appendix \textcolor{red}{G}. As can be inferred from Figures \textcolor{red}{G.I({\it a,b})}, despite the dramatic drop in the number of observations, the difference in the estimated trajectories of cost efficiency scores between the unbalanced (baseline) and balanced panels is minor. 

There are several forces behind this result. First, \cite{Goncharenko2022} find that the closed banks were trying to pool with financially strong banks by mimicking their balance sheets, in particular, by hiding losses on corporate loans and artificially inflating the quality of their assets. The authors also report that there were no ``structural breaks" in the banking system during the active phase of the policy in the sense that, e.g., the ratio of bank total assets to GDP continued rising during the five years of the policy -- from 72\% in the beginning to 91\% in the end. This is because the majority of closed banks were relatively small financial institutions that were not systemically important.\footnote{In our setting, the average size of survived banks equals 161 bn Rubles which is 8 times larger than the average size of the closed banks.} Second, we go further and show that in our setting, the relative differences across the closed and survived banks are not large. As can be inferred from Figure \ref{fig:kdensity}, bank distributions by the ratio of $Revals^-$ to total costs ({\it a}) and the net FX position ({\it b}) exhibit similar patterns for these two groups of banks in terms of means and tails.

\subsection{A counterfactual exercise: Estimating true cost efficiency assuming data on $Revals^-$ is not available} \label{sec:results_counterfact}

Recognizing that our bank-level data on $Revals^-$ is unique and banks in other EMEs may not report it, we propose an alternative strategy that analysts can deploy to obtain reliable estimates of bank cost efficiency. Suppose now that we do not have the bank-level data on $Revals^-$ in the Russian banks' P\&L accounts and we thus cannot compute $CAOC_{it}$ but we still want to construct estimates of bank cost efficiency with the associated FX effects removed. Suppose also that we do observe domestic NER fluctuations and bank-level data on the currency structure of loans and deposits from the banks' balance sheets.\footnote{Banks' holdings of stock and bonds, and other operations, in different currencies are ignored for this exercise. This is to simplify the data requirements to our proposed approach and explore the extent to which accounting for currency structure of only loans and deposits is able to capture the true estimates of cost efficiency.} These two types of data are typically present in EMEs.

In this counterfactual setting, we begin by analyzing the time evolution of FX loans to total assets ratios against the background of the volatility of NER, as measured with the implied volatility of the US Dollar to Ruble exchange rate predicted from a GARCH(1,1)-model. As we show in Figure \textcolor{red}{I.I} in Online Appendix \textcolor{red}{I}, FX loans ratios consistently rise across all groups of (small and large) banks during the crisis periods in 2009 and 2014-2015.\footnote{This is clearly due to corresponding depreciation of Ruble and not due to an increase in borrowers' demand on FX loans, which was actually contracting, given the crises (as in, e.g., Hungary during the same time and for the same reason \citealp{Verner2020}). Of course, FX loans are a part of total assets. The ratio of FX loans to total assets was growing due to the fact that total assets were contracting faster than the Ruble equivalent of the FX loans (recall that banks convert the FX loans using the Ruble exchange rate to report them in their balance sheets).} 
This observation, which is fully consistent with Stylized Fact \#2, gives us a clue on how to approach currency-adjusted cost efficiency without possessing data on $Revals^-$.

We suggest the following two-stage approach: (i) clean out the macro effect of NER fluctuations from bank {\it total} costs; and then (ii) take the residual costs and specify a translog cost function, in which individual exposures to the FX risk are explicitly introduced.\footnote{The same approach can be conducted in a single stage, which we also do and report the results below. The aim of the two-stage approach is to clearly decompose the effects of the macro- and bank-level forces.} Under this FX-adjusted SFA approach, the definition of bank cost efficiency reads as follows: it is the bank's ability to bear the lowest possible costs to produce the same volume of outputs given the market-based factor input prices, the bank managers' risk aversion, and {\it the currency structure of the bank's operations under the observed volatility of domestic NER.}

In the first stage, we run a FE regression of bank operating costs on the lag structure of NER itself or, alternatively, the implied volatility of NER:
\begin{align} \label{eq:SFA_GARCH_1stStage}
    \ln OC_{it} = \alpha_i + \sum^K_{k=0} \beta_k \ln NER_{t-k} +  \varepsilon_{it},
\end{align}
where $K$ is set as either 0, 1, or 2 quarters.\footnote{We also consider deeper lags, up to four quarters, and obtain the same results} We then compute
\begin{align} \label{eq:SFA_GARCH_1stStage_RESID}
    \ln \widehat{OC}_{it} = \ln OC_{it} - \sum^K_{k=0} \hat{\beta}_k \ln NER_{t-k}
\end{align}
and run the second stage as follows:
\begin{align} \label{eq:SFA_GARCH_2mdStage}
    \ln \widehat{OC}_{it} =& f\big(Y_{it}, w_{it}, Risk_{it} \big) + \phi_1 \cdot \ln NER_{it} + \dfrac{1}{2} \phi_2 \cdot  (\ln NER)^2_{it} \\ \nonumber
    & + \sum^{N}_{j=1} \pi_j \cdot \ln Y_{j,it} \times \ln NER_{it} + \sum^{N}_{k=1} \tau_k \cdot \ln w_{k,it} \times \ln NER_{it} + u_{it} + v_{it},
\end{align}
where $f(\cdot)$ is the translog cost function we considered in Equation (\ref{eq:RusBanksCostFrontier}), total loans as previous output is now divided into two as $\ln FX.Loans_{it}$ and $\ln Rub.Loans_{it}$ and
\begin{align} \label{eq:RusBanksCovariates_2Stage}
    u_{it} = \mu(z_{it};\delta) + a_1 \dfrac{FX.Loans_{it}}{TA_{it}} + a_2 \Delta \ln NER_t + a_3 \Big( \dfrac{FX.Loans_{it}}{TA_{it}} \times \Delta_4 \ln NER_t \Big), 
\end{align}
where $\mu(\cdot;\delta)$ is the specification of inefficiency used in Equation (\ref{eq:RusBanksCovariates})  and $\Delta_4 \ln NER_t$ is the four-quarter log-difference of NER capturing the flow nature of  $Revals^-_{it}$, assumed to be unknown.

The estimation results from the first stage appear in Table \ref{tab:Costs_NER_GARCH_main}. We obtain positive and highly significant estimates on both $\ln$ NER and NER Volatility variables (up to the second-quarter lag). These results indicate that following periods of Ruble depreciation, bank costs tend to rise in the subsequent two quarters. When we further plot the time evolution of the fitted costs against the actual operating costs $OC_{it}$, we observe that using the volatility of NER leads to a much better fit compared to the mean NER (Figure \textcolor{red}{H.II} in Online Appendix \textcolor{red}{H}).

\begin{table}[h!] 
\footnotesize
\caption{Estimation results from the first stage} \label{tab:Costs_NER_GARCH_main}
\centering
\begin{tabular}{@{} p{0.25\textwidth} @{} 
>{\centering\arraybackslash}m{0.15\textwidth} @{} 
>{\centering\arraybackslash}m{0.15\textwidth} @{}
>{\centering\arraybackslash}m{0.15\textwidth} @{} 
>{\centering\arraybackslash}m{0.15\textwidth} @{} }
	\toprule[0.5mm]
            &  \multicolumn{4}{c}{Dependent variable $Y_{it}=$ operating costs $OC_{it}$} \\ \cmidrule(r){2-5}  
            & \multicolumn{2}{c}{$X_t=\ln$ NER} & \multicolumn{2}{c}{$X_t=$ NER Volatility,} \\
            &           &                    & \multicolumn{2}{c}{GARCH(1,1)} \\
            \cmidrule(r){2-3} \cmidrule(l){4-5}
	                &  (1) & (2) & (3) & (4)\\ \noalign{\vskip 1mm} \hline \noalign{\vskip 4mm}  

$X_t$              &       1.406***&       1.109***&     0.110***   &    0.056***  \\
                    &     (0.037)   &     (0.031)   &    (0.003)     &   (0.002)     \\ \noalign{\vskip 2mm} 
$X_{t-1}$            &               &       0.407***&               &   0.062***    \\
                    &               &     (0.017)   &               &    (0.002)    \\ \noalign{\vskip 2mm}
$X_{t-2}$            &               &      --0.031   &               &     0.089***   \\
                    &               &     (0.025)   &               &    (0.003)   \\ \noalign{\vskip 4mm}




$N$ Obs    &       47,630   &       47,194   &       47,630   &       47,194   \\
$N$ banks          &        1,236   &        1,226   &        1,226   &        1,226   \\
$R^2_{within}$       &       0.810   &       0.812   &       0.789   &       0.804   \\

\bottomrule[0.5mm] \noalign{\vskip 2mm}
\end{tabular} 
	\begin{minipage}{1\linewidth}
    {\it Note}: Bank controls (the size of a bank's total assets) and quadratic deterministic trend are included but not reported to preserve space. \\ ***, **, * indicate that a coefficient is significant at the 1\%, 5\%, 10\% level, respectively. Robust standard errors are clustered at the bank level and appear in the brackets under the estimated coefficients.   \\		
	\end{minipage}
\end{table}

The estimation results from the second stage appear in Table \ref{tab:Translog_for_Counterfact}. In the translog equation (Panel 1), we see that the estimates on NER, FX loans and their cross-product are statistically significant (at 1\%), meaning that banks with more FX loans experience larger rises in operating costs, especially during periods of Ruble depreciation (rising NER). In the cost inefficiency equation (Panel 2), we also observe positive and significant estimates on the FX loans to total assets ratio and its cross-product with four-quarter changes in NER. This indicates that larger FX loans ratios are associated with spikes in bank cost inefficiency, and the more so during periods of increased NER volatility. 

\begin{table}[h!] 
\footnotesize
\caption{Estimation results from the two-stage}\label{tab:Translog_for_Counterfact}
\centering
\begin{tabular}{@{} p{0.45\textwidth} @{} 
>{\centering\arraybackslash}m{0.15\textwidth} @{} 
>{\centering\arraybackslash}m{0.03\textwidth} @{} 
>{\centering\arraybackslash}m{0.15\textwidth} @{} }
	\toprule[0.5mm]

    \hspace{25mm} NER volatility adjustment & \multicolumn{1}{c}{With lags} & & \multicolumn{1}{c}{Without lags} \\ 
                & ({\it baseline}) && \\ \cmidrule(r){2-2} \cmidrule(l){4-4}
            &  (1) & & (2)  \\ \noalign{\vskip 1mm} \hline \noalign{\vskip 4mm}  

\hspace{0mm}\textit{Panel 1: Cost frontier equation} &  &&  \\ \noalign{\vskip 3mm}

\hspace{2mm} $\ln NER_t$ & --1.805*** && --1.345*** \\ 
                       &  (0.207)   && (0.208)  \\ \noalign{\vskip 2mm}
\hspace{2mm} $\ln^2 NER_t$ & 0.274*** && 0.265*** \\ 
                       &  (0.031)   && (0.031) \\ \noalign{\vskip 2mm}
\hspace{2mm} $\ln NER_t \times \ln FX.Loans$ &   0.043*** && 0.039*** \\ 
                                          &  (0.005)   && (0.005)    \\ \noalign{\vskip 2mm}
\hspace{2mm} $\ln FX.Loans$ &   0.129*** && 0.141*** \\ 
                           &  (0.015)   && (0.015)  \\ \noalign{\vskip 2mm}
\hspace{2mm} $\ln^2 FX.Loans$ &   0.017*** && 0.017*** \\ 
                             &  (0.001)   && (0.001) \\ \noalign{\vskip 3mm}                           
\hspace{2mm} Other cost frontier covariates &  Yes && Yes \\ \noalign{\vskip 2mm}                      
\hspace{2mm} Constant return to scale       &  Yes && Yes \\ \noalign{\vskip 6mm}
                                          
\hspace{0mm}\textit{Panel 2: Inefficiency equation}       \\   \noalign{\vskip 3mm}
\hspace{2mm} $FX.Loans_{it}$ (as \% of $TA_{it}$) & 0.026*** && 0.029*** \\
                                      & (0.002)  && (0.003) \\ \noalign{\vskip 2mm}
\hspace{2mm} $\Delta_4 \ln NER_t$ & --0.124  && 1.075*** \\
                                         & (0.193)  && (0.237)  \\ \noalign{\vskip 2mm} 
\hspace{2mm} $FX.Loans_{it}$ (as \% of $TA_{it}$) $\times$ $\Delta_4 \ln NER_t$ & 0.018*** && 0.009* \\
                                                   & (0.005)  && (0.006) \\ \noalign{\vskip 2mm} 
\hspace{2mm} Other mean inefficiency covariates &  Yes && Yes \\ \noalign{\vskip 2mm}

$N$ Obs   &  19,240 && 19,240 \\
$N$ Banks                   &  805    && 805    \\
log likelihood          &  --2,980.4 && --3,245.0 \\
Convergence achieved    &  Yes && Yes  \\
\bottomrule[0.5mm] \noalign{\vskip 2mm}
\end{tabular} 
	\begin{minipage}{1\linewidth}
    {\it Note}: ***, **, * indicate that a coefficient is significant at the 1\%, 5\%, 10\% level, respectively. Standard errors are clustered at the bank level and appear in the brackets under the estimated coefficients.   \\
	\end{minipage}
\end{table}

Finally, we re-compute the cost efficiency scores and compare them with the original estimates. We find that the mean value of the two-stage cost efficiency scores is 78.7\% (Table \ref{tab:desc_stats_2Stage}). This is 21 pp larger than the original estimate obtained without dropping  $Revals^-$ (57.8\%), though still 11 pp lower than the original estimate obtained after dropping $Revals^-$ (89.9\%). There is still some difference, which is likely due to the fact that we omit other FX operations from the cost frontier estimation. However, our results clearly indicate that the proposed two-stage approach is capable of reducing the distance between the available and true measures of cost efficiency -- by about two-thirds in the case of Russian data. Moreover, if we repeat the same exercise with only the first stage (i.e., eliminating the macro force of NER) or only the second stage (i.e., removing the bank-specific exposures to the FX risk), we would obtain mean cost efficiency scores of 67.1\% and 76.8\%, respectively. This shows the relative importance of the two underlying forces and that macro effects are dominated by bank-specific considerations. Running the two stages in one produces a mean cost efficiency score which is only 1 p.p. lower than in our two-stage approach and is statistically indistinguishable from it. 

\begin{table}[h!] 
\footnotesize
\caption{Comparison of true and counter-factual estimates of cost efficiency scores}\label{tab:desc_stats_2Stage}
\centering
\begin{tabular}{@{} p{0.5\textwidth} @{} 
>{\centering\arraybackslash}m{0.08\textwidth} @{} 
>{\centering\arraybackslash}m{0.08\textwidth} @{}
>{\centering\arraybackslash}m{0.08\textwidth} @{}
>{\centering\arraybackslash}m{0.08\textwidth} @{}
>{\centering\arraybackslash}m{0.08\textwidth} @{} }
\toprule[0.5mm]
	                      
	                   & Obs & Mean & SD & Min & Max \\  \noalign{\vskip 1mm} 
	                   \cmidrule(r){2-6} 
	                   & (1) & (2)       & (3)            & (4)   & (5)                  \\ \noalign{\vskip 2mm} \hline \noalign{\vskip 2mm} 

\multicolumn{6}{l}{\hspace{-2mm} {\it Panel 1: Baseline (true) estimates: }} \\ \noalign{\vskip 2mm}
\hspace{4mm} $Revals^-$ dropped &  19,240  &  89.9  &  8.8  &  9.0  &  99.6    \\
\hspace{4mm} $Revals^-$ kept    &  19,240  &  57.8  &  23.5  &  0.6  &  99.2 \\
\noalign{\vskip 4mm} 

\multicolumn{6}{l}{\hspace{-2mm} {\it Panel 2: Estimates from the two-stage approach (counter-factual):} } \\ \noalign{\vskip 2mm}
\hspace{4mm} Both stages & 19,240  &  78.7  &  15.7  &  1.5  &  97.2  \\
\hspace{4mm} Only first stage & 19,240  &  67.1  &  20.5  &  0.8  &  96.5  \\
\hspace{4mm} Only second stage & 19,240  &  76.8  &  17.5  &  1.5  &  98.1  \\
\hspace{4mm} Two stages in one stage & 19,240  &  77.2  &  17.3  &  1.4  &  98.0  \\
 \noalign{\vskip 2mm}
\bottomrule[0.5mm] \noalign{\vskip 2mm}
\end{tabular}
\end{table}

We also analyze the time evolution of the two-stage cost efficiency scores in comparison to the original estimates.\footnote{For the sake of space, we report the results and their description in Online Appendix \textcolor{red}{H}.} One important outcome from this analysis is that our two-stage approach reduces substantially the pro-cyclicality of the estimated cost efficiency scores, which makes them much closer to the true estimates when $Revals^-$ are available and dropped explicitly from the operating costs. Overall, we propose an adaptation of an otherwise standard stochastic frontier analysis to measure bank cost efficiency when banks operate in dollarized EMEs characterized by increased NER volatility. 

\section{Cross-sectional heterogeneity} \label{sec:channels}
\subsection{$Revals^-$ and cost efficiency: potential channels}
Given the crucial role of $Revals^-$ in shaping bank performance, we now explore the empirical relationship between $Revals^-$ and the combination of bank-specific factors and the Ruble's exchange rate in the cross-section of banks. From Stylized Fact \#4, a natural choice for exploring the cross-sectional variation would be considering a bank's net FX position (FX assets net of FX liabilities). Rising positive (negative) net FX positions should predict greater $Revals^-$ during times of Ruble appreciation (depreciation). 
Further, 
it may be the case that larger or faster-growing banks may reduce the adverse ``effect" of their net FX positions on $Revals^-$ through diversification of their FX assets and liabilities across countries and/or through the use of forward currency contracts \citep{Ippolito2002}. 

\begin{table}[h!] 
\footnotesize
\caption{Heterogeneous relationships between $Revals^-$ and net FX positions of banks}\label{tab:Channels}
\centering
\begin{tabular}{@{} p{0.4\textwidth} @{} 
>{\centering\arraybackslash}m{0.12\textwidth} @{} 
>{\centering\arraybackslash}m{0.12\textwidth} @{} 
>{\centering\arraybackslash}m{0.12\textwidth} @{} 
>{\centering\arraybackslash}m{0.12\textwidth} @{} }
	\toprule[0.5mm]
\hspace{38mm} $Y_{it}$:            & \multicolumn{4}{c}{Dependent variable = $Revals^-_{it}$ to $TC_{it}$ ratio} \\ \noalign{\vskip 2mm} \cmidrule(r){2-5}
\hspace{38mm} $X_{it}$:      & 
            $\ln TA_{it}$ & $\dfrac{EQ_{it}}{TA_{it}}$ & $\dfrac{LIQ_{it}}{TA_{it}}$ & $\Delta_4 \ln TA_{it}$  
	          \\ \noalign{\vskip 2mm} \cmidrule(r){2-5}
            &  (1) & (2) & (3) & (4)    \\ \noalign{\vskip 1mm} \hline \noalign{\vskip 4mm}  

Positive net FX position$_{it-1}$ &       0.082***&       0.040** &       0.057***&       0.060***   \\
                    &     (0.020)   &     (0.018)   &     (0.013)   &     (0.013)      \\ \noalign{\vskip 1mm}
Negative net FX position$_{it-1}$ &      --0.220***&      --0.123***&      --0.159***&      --0.160***  \\
                    &     (0.020)   &     (0.019)   &     (0.016)   &     (0.014)      \\  \noalign{\vskip 4mm}

Positive net FX position$_{it-1}$ $\times$ $X_{it-1}$
		    &      --0.022*  &    0.016    &    0.004  &   --0.024       \\ 
                    &     (0.013)   &     (0.015)     &  (0.020)     &    (0.052)     \\ \noalign{\vskip 1mm}
Negative net FX position$_{it-1}$ $\times$ $X_{it-1}$
		    &       0.068***&    --0.035**   &    0.000    &    0.015              \\
                    &     (0.013)   &    (0.017)  &    (0.012)   &    (0.031)      \\ \noalign{\vskip 4mm}

                    
Bank controls, $X_{it-1}$ & Yes & Yes   & Yes & Yes \\
Bank FEs  & Yes & Yes   & Yes & Yes \\
Quarter FEs & Yes & Yes & Yes   & Yes \\ \noalign{\vskip 3mm}
Obs   & 29,881 & 29,881   & 29,881 & 29,881  \\
Banks & 993 & 993 & 993   & 993 \\
$R^2_{within}$ &  0.506   &       0.502   &       0.501   &       0.501      \\

\bottomrule[0.5mm] \noalign{\vskip 2mm}
\end{tabular} 
	\begin{minipage}{1\linewidth}
    {\it Note}: The table reports two-way FEs regression of the $Revals^-$ to total cost ratio on (i) positive and negative net FX positions, (ii) their cross-products with bank size ($\ln TA$), capitalization ($EQ/TA$), liquidity ($LIQ/TA$, where $LIQ$ includes cash holdings and reserves), and annual growth rate of total assets ($\Delta_4 \ln TA$), and (iii) all components of the cross-products. 
    All variables were divided by their corresponding standard deviations.
    \\ ***, **, * indicate that a coefficient is significant at the 1\%, 5\%, 10\% level, respectively. Standard errors are clustered at the bank level and appear in the brackets under the estimated coefficients.   \\		
	\end{minipage}
\end{table}

We incorporate these ideas in the following panel FE regression model:
\begin{align}
    \dfrac{Revals^-_{it}}{TC_{it}} & = \alpha_i + \beta_t + \gamma_1 \cdot Net\,FX^-_{it-1} +  \gamma_2 \cdot Net\,FX^+_{it-1} \\ \nonumber
    & + \gamma_3 \cdot  \Big( Net\,FX^-_{it-1} \times X_{k,it-1} \Big) + \gamma_4 \cdot  \Big( Net\,FX^+_{it-1}\times X_{k,it-1}\Big)  + \mathbf{X}_{it-1} \Psi' + \epsilon_{it},
\end{align}
where $k=1,\ldots,4$ and $X_{it}$ is bank size ($\ln TA$), capitalization ($EQ/TA$), liquidity ($LIQ/TA$, where $LIQ$ includes cash holdings and reserves), and annual growth rate of total assets ($\Delta_4 \ln TA$). All RHS variables were centered around zero to facilitate interpretation.

The estimation results appear in Table \ref{tab:Channels}. First, we obtain positive and significant (almost always at the 1\% level) estimates on the $Net \, FX^+_{it-1}$ variable and negative and significant (at the 1\% level) estimates on the $Net\,FX^-_{it-1}$ variable across all specifications. These estimates imply that larger deviations of net FX position from zero in either direction are associated with larger $Revals^-$. Moreover, we obtain substantial asymmetry: a one standard deviation change in the negative FX position raises the $Revals^-$ contribution to total costs by 0.12 to 0.22 of the $Revals^-$'s standard deviation, whereas the corresponding economic effect of a one standard deviation change in the positive net FX position is 3 times lower in magnitude, other things equal. 

Second, we obtain a negative and marginally significant estimate on the cross-product of bank size ($\ln TA_{it}$) and $Net\,FX^+$ and a positive and highly significant estimate on the cross-product of bank size ($\ln TA_{it}$) and $Net\,FX^-$. This means that a larger bank may reduce the adverse ``effect" of both negative and positive net FX positions on the bank's $Revals^-$ and thus total costs, as hypothesized above. 

Third, we also obtain a negative and significant coefficient on the cross-product of $Net\,FX^-$ and bank capital ($EQ_{it}/TA_{it}$). This suggests that more leveraged banks are also able to limit the ``effect" of  negative net FX positions on their $Revals^-$. Other potential channels of heterogeneity such as bank liquidity ($LIQ_{it}/TA_{it}$) and business growth ($\Delta_4\ln TA_{it}$) do not exhibit statistical significance. It is thus clear that $Revals^-$ are mostly generated by negative net FX positions of relatively small and/or more capitalized banks. 

Overall, we see that negative net FX positions raise $Revals^-$ which, if kept inside the operating costs, lead to spurious conclusions such as the pro-cyclicality of managerial ability to run banks or pervasive inefficiency of foreign-owned banks. 

\subsection{Tail dependence in cost efficiency rankings}
In Section \ref{sec:scores}, we showed that dropping $Revals^-$ from the LHS of the translog cost function while keeping the RHS unchanged mechanically raises the resultant cost efficiency scores (recall Figure \ref{fig:SFA_TimeEvol}). However, what is much more important is whether dropping $Revals^-$ changes banks' rankings in terms of cost efficiency scores.\footnote{Clearly, if the banks' ranking preserves, $Revals^-$ would not affect answering the question ``Who is more efficient and who is less?" and the importance of $Revals^-$ would diminish.} As our computations show, the rankings indeed change and there is no clear mapping between the two rankings, i.e., with $Revals^-$ kept or dropped (Figures \textcolor{red}{I.I-II} in Online Appendix \textcolor{red}{I}).

\begin{figure}[h!]
    \centering
    \subfigure[2005Q1: Beginning of the sample period]{
    \includegraphics[width=0.45\textwidth]{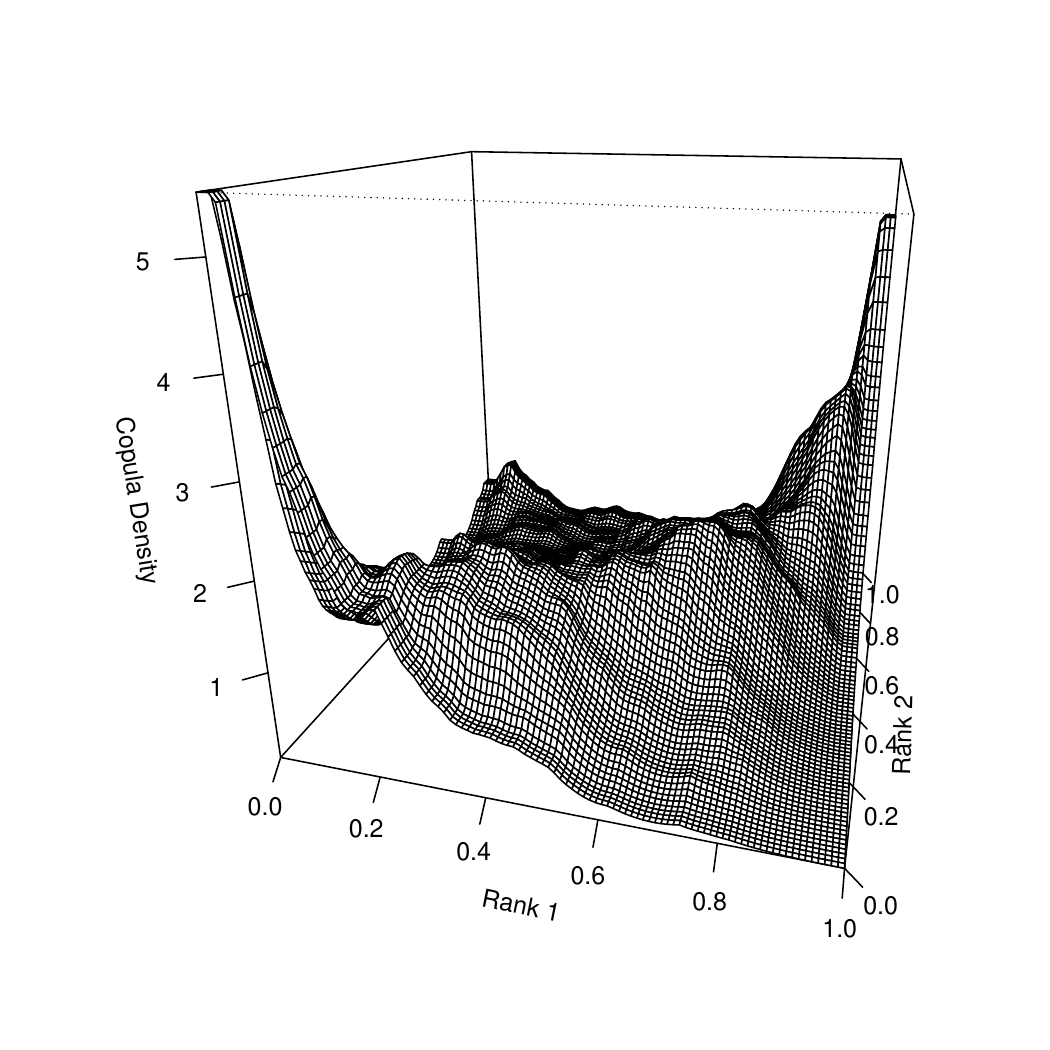}}
    \subfigure[2009Q2: The Global Financial Crisis]{
    \includegraphics[width=0.45\textwidth]{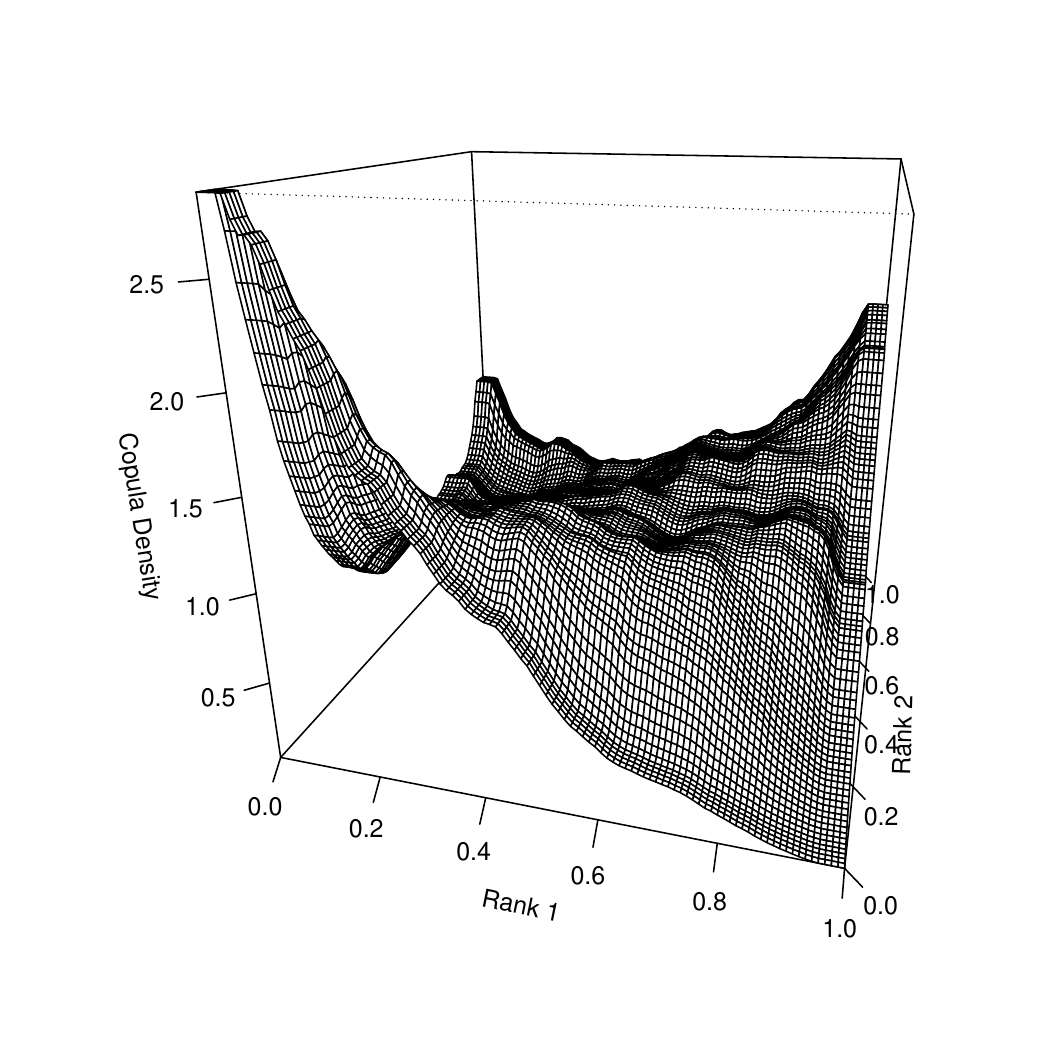}}
    
        \vspace*{\floatsep}
        
    \centering
    \subfigure[2014Q2: First sanctions against Russia]{
    \includegraphics[width=0.45\textwidth]{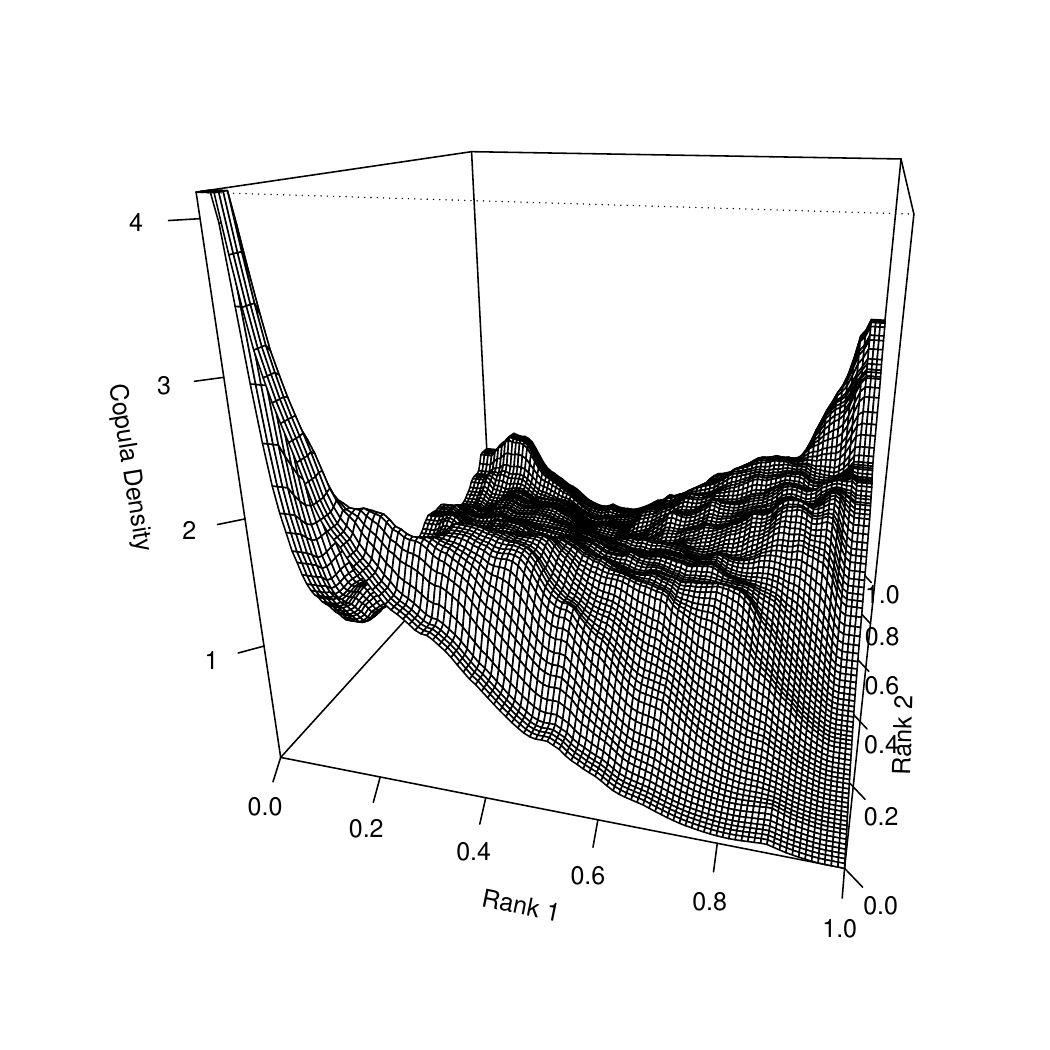}}
    \subfigure[2020Q2: The end of the sample period]{
    \includegraphics[width=0.45\textwidth]{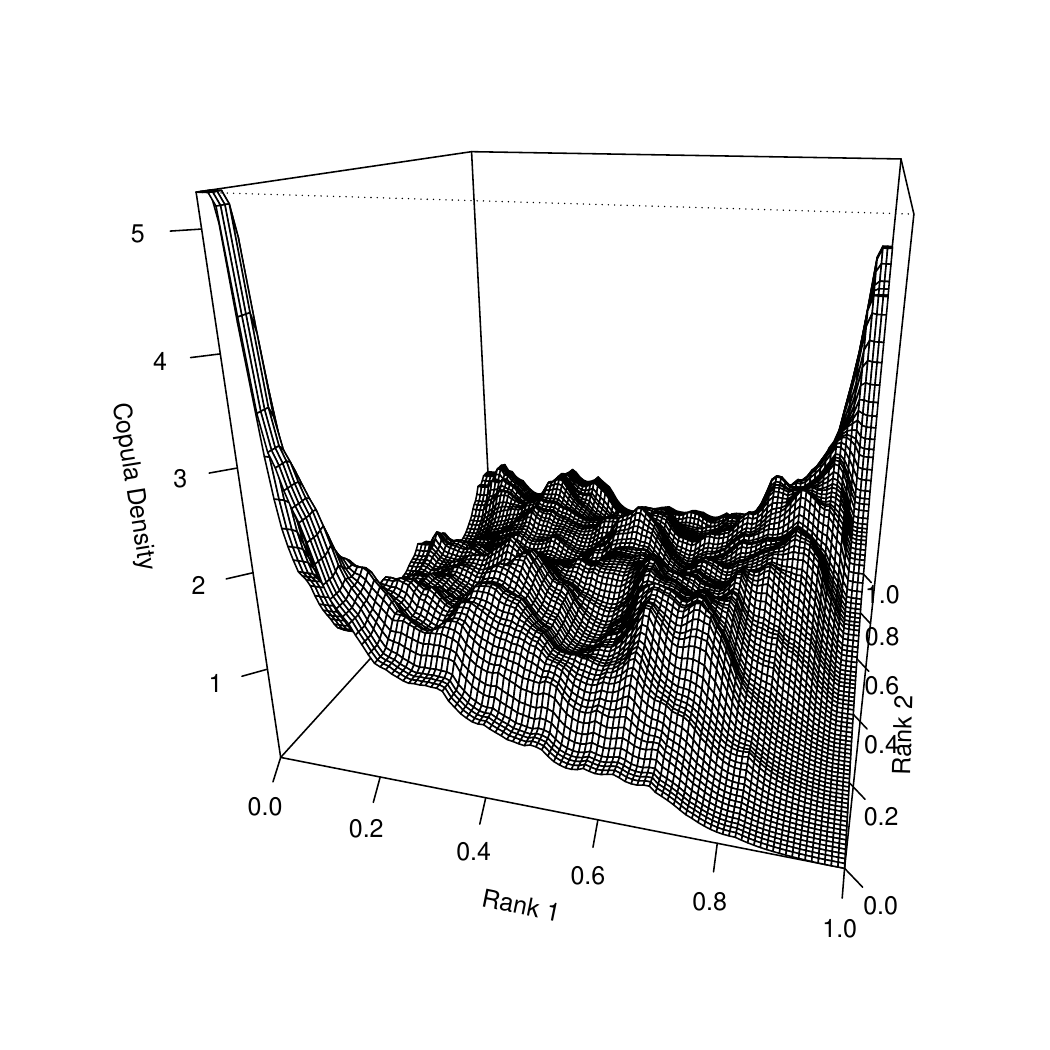}}
    
    \begin{minipage}{1\linewidth}
    \vspace{3mm}
    		\footnotesize	
    		{\it Note}: The results are based on the {\tt npcopula} routine in the {\tt R} package {\tt np} \citep{Racine2008}. 
    	\end{minipage}
    \captionsetup{justification=centering,margin=2cm}
    \caption{Smoothed copulas on the rankings computed based on \\ the cost efficiency scores with $Revals^-$ kept \textbf{(Rank 1)} \\ and dropped \textbf{(Rank 2)}, for specific quarters.}
    \label{fig:RusBanks_COPULAS}
\end{figure}

To better assess dependence amongst the two cost efficiency rankings we deploy a smoothed copula density approach \citep{Racine15}. We estimate these smoothed densities using data-driven bandwidth selection and report the results in Figure \ref{fig:RusBanks_COPULAS}, which provides a snapshot of the results for four selected quarters: 2005 Q1 (start of the sample period, {\it a}), 2009 Q2 (the most acute phase of GFC, {\it b}), 2014 Q2 (first sanctions, {\it c}), and 2020 Q2 (end of the sample period, {\it d}).\footnote{The rest of the results are not disclosed for the sake of space and are available upon request.} We obtain no strong dependence between the two rankings for a majority of banks in the middle of the distribution across all quarters. However, for very efficient and very inefficient banks we uncover evidence of {\it extreme tail dependence} in the estimated cost efficiency rankings, represented by an accumulation of mass in the opposite corners of estimated copulas. In particular, the banks that are the most efficient when $Revals^-$ are kept are also found to be among the most efficient when $Revals^-$ are dropped (the (1,1) corner). The same holds for the most inefficient banks (the (0,0) corner). These patterns are persistent in time.


We also observe a substantial amount of probability mass in the $(0,1)$ corner across all quarters. Some banks that are heavily inefficient when assessed with the inclusion of $Revals^-$ become highly efficient once $Revals^-$ are dropped. Further analysis shows that these are the {\it Big-4} state-owned banks and a bulk of the foreign banks. This is consistent with our estimation results on foreign banks discussed in Section \ref{sec:CostEfficiency_Results_Regression}. Notably, these two groups have the largest exposures to foreign currencies in their operations.

Finally, we observe virtually no banks in the $(1,0)$ area of the ranking space. This suggests that if banks are inefficient even when $Revals^-$ are kept, they would not improve their relative ranking when $Revals^-$ are dropped.  

\section{
Broader implications of currency revaluations} \label{sec:Broad}

In this section, we address more fundamental economic issues related to ignoring the adverse effects of $Revals^-$. The first is how this ignorance affects the relationship between the efficiency of financial intermediation and the {\it credit market structure}. The second is {\it financial stability}: how banks can suppress the contribution of $Revals^-$ to their overall risk exposures.

\subsection{Credit market structure} \label{sec:ESHvsQLH}

In general, the {\it Efficient Structure Hypothesis} (ESH) states that more efficient banks gain higher market shares by replacing less efficient rivals \citep{Demsetz73}. A valid mechanism is that  more cost efficient banks can extend loans to the real economy at lower interest rates \citep{Gambacorta2008}.\footnote{This price mechanism is of utmost importance for EMEs like Russia, given that their larger firms tend to borrow cheaper from abroad \citep{Bruno17}, thus reducing domestic loan demand and the growth potential of local banks, and smaller firms tend to encounter difficulties in obtaining loans, especially in the localities with strong presence of foreign banks \citep{Hasan2017}.} In turn, the {\it Quiet Life Hypothesis} (QLH) implies that the ability of more efficient banks to replace less efficient rivals may be weak in the presence of a monopoliest, or a set of colluding oligopolists \citep{Berger98}. Previous studies that have tested ESH against QLH did not focus on the role of $Revals^-$ \citep[see, e.g.,][]{Delis09, Koetter12, Kumbhakar14a, Huang18}. Since ignoring $Revals^-$ contaminates the estimates of bank cost efficiency, the outcome of testing between the two hypotheses is unclear.


More formally, we test ESH against QLH using 
the following two-way FEs regression:
\begin{align} \label{eq:RusBanksSFALoans}
    \ln MS_{\jmath,it} = \alpha_{\jmath,i} + \beta_{\jmath,t} + \gamma_{\jmath} \ln \widehat {CE}^{(\ell)}_{it} + \mathbf{X}_{it} \Psi' + \varepsilon^{(\ell)}_{\jmath,it},
\end{align}
where $MS_{\jmath,it}$ is the $\jmath^{th}$ market share of bank $i$ in quarter $t$, where $\jmath=1$ corresponds to the corporate credit market and $\jmath=2$ to the household credit market. We separate the credit market into these two segments because of substantial differences in the currency composition of loans.\footnote{As of 2019, credit in foreign currencies constituted only about 2\% of total credit to households while being much larger in the case of firms, nearly 25\%.}

If $\gamma_{\jmath} > 0$, we reject QLH and claim our empirical evidence favors ESH. If $\gamma_{\jmath} = 0$, we conclude the opposite. The case of $\gamma_{\jmath} < 0$ is also possible and, as explored by \cite{Koetter12}, would indicate that gaining greater market shares requires bank managers to bear non-market-based costs to diminish bank competition (corruption costs, signalling costs, regulatory costs, etc.).

To understand the role of $Revals^-$ in the outcome of testing ESH against QLH, let us first note that we can write $CE^{(1)}_{it}=CE^{(2)}_{it} \times \xi_{it}$, where  $\xi_{it}\in[1,1/CE^{(2)}_{it}]$ is a mapping parameter that captures the 
cost efficiency gap (Section \ref{sec:scores}). We can then rewrite Equation~\eqref{eq:RusBanksSFALoans} as:
\begin{align} \label{eq:RusBanksSFALoans2}
    \ln MS_{\jmath,it} & = \alpha_{\jmath,i} + \beta_{\jmath,t} + \gamma_{\jmath} \ln \widehat {CE}^{(1)}_{it} + \mathbf{X}_{it} \Psi' + \varepsilon^{(1)}_{\jmath,it} \notag\\
    &= \alpha_{\jmath,i} + \beta_{\jmath,t} + \gamma_{\jmath} \ln \widehat {CE}^{(2)}_{it} + \gamma_{\jmath} \xi_{it} + \mathbf{X}_{it} \Psi' + \varepsilon^{(1)}_{\jmath,it}.
\end{align}
It is clear that if $\xi_{it}$ varies with any of the control variables in $\mathbf{X}_{it}$ or with $CE^{(2)}_{it}$ and it is omitted then we have an omitted variable bias in the estimation of $\gamma_{\jmath}$. Thus, attempting to discern the impact of estimated cost efficiency on market share will be dependent upon which measure is used. 
In economies with stable currencies ($\xi_{it} \rightarrow 1$), banks can exhibit little differentiation in $\widehat {CE}^{(1)}_{it}$ and $\widehat {CE}^{(2)}_{it}$, and the expectation is that $\gamma_{\jmath}$ adequately captures evidence for or against  QLH. However, in economies with large fluctuations in the exchange rate ($\xi_{it}>1$), it is likely that the measurement of cost efficiency will have an impact on assessing  QLH. 

The estimation results appear in Table \ref{tab:results_SFA_and_Lending_main}. We begin with an OLS regression and then augment the estimation procedure by applying GMM with ${CE}^{(\ell)}_{it-1}$ used as an instrument for ${CE}^{(\ell)}_{it}$ (exact identification).\footnote{The results do not change qualitatively if we use more lags.} 
GMM addresses the simultaneity concern (namely, greater market shares are likely to affect cost efficiency via, e.g., improving economies of scale). Using past values of efficiency as instruments isolates a part of the variation in ${CE}^{(\ell)}_{it}$ which is driven by its own inertia rather than the current changes in market shares $ MS^{(\jmath)}_{it}$. Further, given that the Russian banking system is populated with a few very large banks (mostly government-connected, \citealp{Goncharenko2022}) and a myriad of very small banks, we also show how the estimation results vary if we drop the  {\it smallest} 25\% of banks and the  {\it largest} 25\% of banks in terms of total assets. Online Appendix \textcolor{red}{J} contains the underlying descriptive statistics. We address the issue of generated regressors by bootstrapping (1,000 draws) the standard errors of the estimated coefficients.

{\it Corporate credit market ($\jmath =1$)}. In the OLS regression reported in columns (1) and (2), we obtain a positive and highly significant estimate on $\ln CE^{(\ell)}_{it}$ when $Revals^-$ are dropped ($\ell =1$) and an insignificant estimate when $Revals^-$ are kept ($\ell =2$). This is a clear indication that $Revals^-$ may obscure the relationship between efficiency and market shares. Further GMM refinements of the two estimates reported in columns (3) and (4) deliver positive and highly significant estimates in both cases, however, the underlying economic effects differ substantially. In response to a one standard deviation change in $CE^{(\ell)}_{it}$, corporate credit grows by an additional 3.7 pp when $Revals^-$ are dropped (column (3)) and by only 2.1 pp when $Revals^-$ are kept (column (4)). Given that we control for time FEs which, among other things, absorb changes in the aggregate prices in the economy, the difference in the two estimates -- 1.6 pp -- is stark and has substantial implications for growth in the economy. 

Though evaluating the effects of credit expansions on economic growth goes beyond the scope of this paper, we can well use the estimated elasticities of (real) GDP to bank credit from existing studies. For example, as \cite{Gambetti2017} establish, these elasticities for advanced economies reside around 1. Clearly, in EMEs like Russia, these elasticities are likely to be larger reflecting a lower satiation of domestic credit markets and greater propensity to borrow. Thus, assuming unit elasticity would give a conservative estimate of the real implications of ignoring $Revals^-$. Recall that the share of corporate loans in the total loans of the Russian banking system is roughly 0.5. Combining the estimated difference of 1.6 pp, the unit elasticity of GDP to bank credit, and the 0.5 share of corporate credit in the system, we conclude that ignoring $Revals^-$ results in at least 0.8 pp of ``lost'' GDP growth.  


Furthermore, in columns (5) and (6) we squeeze the sample by removing the {\it largest 25\% of banks} (in terms of total assets) and re-estimate the same model by GMM. We again obtain positive and statistically significant estimates, regardless of whether we drop or keep $Revals^-$, as in the full sample. Quantitatively, the economic effects are now larger by 0.6--1.0 pp than in the full sample, meaning that smaller banks grow faster than the largest banks when they improve cost efficiency.

However, when we move to columns (7) and (8), where we drop the {\it 25\% smallest banks}, we find that the economic effects for the corporate credit market decline substantially, by 1.4--1.7 pp, compared to the full sample. Moreover, the estimated coefficient is positive and statistically significant only when we drop $Revals^-$ ($\ell=1$). When $Revals^-$ are kept, the estimated coefficient turns insignificant ($\ell=2$). Hence, if one analyzes larger banks in an economy with unstable local currency, she could misleadingly conclude that the market for corporate credit is not competitive (QLH dominates), whereas in reality it is competitive (ESH prevails) but the degree of competition is contaminated by the accumulated $Revals^-$. 

\begin{landscape}
\begin{table}
\footnotesize
\caption{$Revals^-$, cost efficiency and bank lending to real economy: estimation results}\label{tab:results_SFA_and_Lending_main}
\centering
\begin{tabular}{@{} p{0.45\textwidth} @{} 
>{\centering\arraybackslash}m{0.11\textwidth} @{} 
>{\centering\arraybackslash}m{0.11\textwidth} @{}
>{\centering\arraybackslash}m{0.11\textwidth} @{} 
>{\centering\arraybackslash}m{0.11\textwidth} @{} 
>{\centering\arraybackslash}m{0.05\textwidth} @{} 
>{\centering\arraybackslash}m{0.11\textwidth} @{} 
>{\centering\arraybackslash}m{0.11\textwidth} @{}
>{\centering\arraybackslash}m{0.11\textwidth} @{} 
>{\centering\arraybackslash}m{0.11\textwidth} @{}}
	\toprule[0.5mm]
    
\hspace{40mm} Level:  & \multicolumn{4}{c}{Full sample: $Q_1:Q_4$}  & & \multicolumn{2}{c}{Subsample: $Q_1:Q_3$} & \multicolumn{2}{c}{Subsample: $Q_2:Q_4$ } \\
                & & & & & & \multicolumn{2}{c}{w/out largest banks} & \multicolumn{2}{c}{w/out smallest banks} \\ \cmidrule(l){2-5} \cmidrule(l){7-8} \cmidrule(l){9-10}
\hspace{40mm} Estimator:  & \multicolumn{2}{c}{OLS} & \multicolumn{2}{c}{GMM} & & \multicolumn{2}{c}{GMM} & \multicolumn{2}{c}{GMM} 
	   \\ \cmidrule(r){2-3} \cmidrule(l){4-5} \cmidrule(r){7-8} \cmidrule(l){9-10} 
	   \hspace{40mm} $Revals^-_{it}$: & Dropped & Kept & Dropped & Kept & & Dropped & Kept & Dropped & Kept  \\ 
        \cmidrule(l){2-10}    &  (1) & (2) & (3) & (4) & & (5) & (6) & (7) & (8)   \\ \noalign{\vskip 1mm} \hline \noalign{\vskip 2mm}  

\multicolumn{10}{l}{\hspace{-2mm} {\it Dependent variable $Y_{it} = \text{ln} (MS^{(1)}_{it})$, \% (corporate credit market)}} \\ \noalign{\vskip 3mm}
\hspace{2mm} $ X_{it} = \text{ln} (CE^{(\ell)}_{it})$ 
                    &       0.424***&     0.063    &       0.503***&     0.107***   & &      0.709***&     0.250***   &       0.326***&    0.024   \\
                    &     (0.066)   &    (0.058)    &     (0.084)   &  (0.025)       & &    (0.127)   &  (0.036)    &     (0.098)   &   (0.028)   \\ \noalign{\vskip 4mm}
                    
\hspace{4mm} Absolute impact, p.p. of $Y_{it}$ 
                    &      3.04*** &    1.65     &    3.73***  &    2.11***    &&      4.89***  &   4.30***   &    2.35*** &  0.44    \\
                    &     (0.87)   &    (1.23)    &     (0.64)   &     (0.48)    & &   (0.88)   &  (0.64)   &    (0.68)   &    (0.52)   \\ \noalign{\vskip 2mm}
                    

$N$ obs   &       37,908   &       37,910   &       37,538   &       37,542   & &       27,563   &       27,566   &       30,064   &       30,067    \\
$N$ banks       &        1,106   &        1,106   &     1,073   &    1,073   & &        898   &    898   &    932      &   932     \\
$R^2$ (adj.) &       0.399   &       0.397   &       0.401   &       0.399   &  &       0.405   &       0.404   &       0.343   &       0.341  \\
\noalign{\vskip 7mm}

\multicolumn{10}{l}{\hspace{-2mm} {\it Dependent variable $Y_{it} = \text{ln} (MS^{(2)}_{it})$, \% (household credit market)}} \\ \noalign{\vskip 3mm}
\hspace{2mm} $ X_{it} = \text{ln} (CE^{(\ell)}_{it})$ 
                    &       0.640***&     0.272***   &       0.850***&    0.308***   &  &     0.796***&   0.260***    &       0.882***&    0.282***   \\
                    &     (0.070)   &     (0.027)   &     (0.089)   &  (0.031)     & &    (0.103)   &    (0.034)     &     (0.099)   &     (0.032)   \\ \noalign{\vskip 4mm} 

\hspace{4mm} Absolute impact, p.p. of $Y_{it}$ 
                    &      4.56*** &    4.81***     &    6.05***  &    5.45***    &&      5.43***  &   4.48***   &   6.27***&   5.16***    \\
                    &     (1.16)   &    (1.52)    &     (0.63)   &     (0.53)    & &   (0.73)   &  (0.58)   &    (0.69)   &    (0.61)   \\ \noalign{\vskip 2mm}

$N$ Obs   &       37,715   &       37,717   &       37,356   &       37,360   & &       27,556   &       27,559   &       29,844   &       29,847   \\
$N$ banks    &        1,094   &        1,094   &   1,061  &   1,061    & &       887 & 887  & 922 & 922  \\
$R^2$ (adj.) &       0.383   &       0.383   &       0.383   &       0.384   & &       0.313   &       0.314   &       0.365   &       0.364   \\
\bottomrule[0.5mm] \noalign{\vskip 2mm}
\end{tabular} 
	\begin{minipage}{1\linewidth}
    {\it Note}: All regressions contain the full set of bank-specific and macro control variables, bank and quarter FEs. $GMM$ uses exact identification (instrument = first lag of respective LHS-variable). \\ 
    ***, **, * indicate that a coefficient is significant at the 1\%, 5\%, 10\% level, respectively. Standard errors are bootstrapped (replications = 1,000, with repetitions) and appear in the brackets under the estimated coefficients.   \\		
	\end{minipage}
\end{table}
\end{landscape}

{\it Household credit market ($\jmath=2$)}. In the full sample, the OLS and GMM regressions reported in columns (1)--(4) also deliver strong support for ESH against QLH, regardless of $Revals^-$. The latter is expected, given that the market for household credit almost fully operates in Rubles, as we mentioned above. In terms of economic effects, we obtain that a one standard deviation rise in the estimated cost efficiency raises household credit by an additional 6.1 pp. This estimate is two to three times larger than its analog for the corporate credit market, thus implying that the household credit market is much more competitive. In columns (5)--(8), where we vary the sample by dropping either the smallest or the largest banks, we obtain significant estimates that do not change meaningfully with respect to the full sample.

Overall, our results for Russian banks indicate that more cost efficient banks accelerate both corporate and household credit growth and earn greater market shares in the banking system. These results are in line with those obtained for Latin American banks by \cite{Tabak12} and globally by \cite{Delis12}. However, we stress the importance of the currency structure of total costs in obtaining these results. Once the currency revaluations on larger banks' balance sheets are ignored, one could misleadingly conclude that the credit markets in EMEs like Russia are inefficient.

\subsection{Financial stability } \label{sec:stability}
\subsubsection{\it Do $Revals^-$ matter for the financial stability? }
As we discuss in the Introduction, currency mismatches have been consistently rising across EMEs after the Global Financial Crisis of 2007--2009 \citep{Kuruc2017,Bruno20} exposing ``mismatched" banks to the risk of additional losses during periods of exchange rate instability.\footnote{This raises the question why at all then banks would be willing to subject themselves to currency mismatch. As has been vividly shown by \cite{Ranciere2010}, engaging in currency mismatches can relax borrowing constraints -- especially for smaller firms and especially during the tranquil times in the economy. Of course, the price the firms, and eventually their banks, have to pay is a much deeper fall when a crisis hits compared to the borrowers with perfectly matched currency structure of the balance sheet.} Losses stemming from negative currency revaluations directly reduce bank profits and thus likely increase the probability of bank distress which, in turn, may then propagate to the real sector of the economy resulting in greater household defaults and capital misallocation \citep{Verner2020,Blattner2023}. 

However, as our copula analysis has shown in the previous section, there is a large cross-sectional variation in how effectively banks manage their FX risk exposures (winners and losers). More broadly, in the case of a perfect hedge (FX assets equal FX liabilities), the losses can be fully offset by the income originating from the other side of the balance sheet, thus eliminating the propagation of rising $Revals^-$ to the banks' exposures to financial distress. But in other cases, the losses can exceed the income, thus creating a direct threat to financial stability and also an indirect ``credit risk" threat through the monetary authority's greater willingness to sustain higher interest rates in the economy \citep{Goldstein2004}.\footnote{Reducing the regulatory interest rate in an EME with currency mismatch would lead to capital outflow which, in turn, would depreciate the local currency and raise the losses of mismatched banks.} We therefore need to understand the patterns of the relationship between negative currency revaluations and banks' exposures to distress, and how this depends on hedging of the FX risk. 

We apply the widely adopted accounting-based measure of bank stability, namely, the Z-score of the distance to default \citep{Beck2013,Bertay2013,Bouvatier2023}. Z-score shows the number of standard deviations ($\sigma(ROA)$) of bank profitability ($ROA$), by which the ROA should fall to fully destroy bank capital. We also complement our analysis with an ex-post measure of credit risk, as reflected in bank non-performing loans ($NPL$).

Our idea is that all banks can be divided into those having relatively large or relatively small portions of operations abroad -- we will call them international and domestic banks, for expositional brevity. Given the instability of exchange rates in countries like Russia, we hypothesize that having operations abroad can serve as a tool to smooth the fluctuations in costs and income, thus offsetting (in part or in full) the negative effects of $Revals^-$ on bank Z-scores and NPLs. We test this hypothesis using the following panel regression:
\begin{align} \label{eq:z_score}
    Risk_{it} =& \, \beta \cdot Revals^-_{it-4} + \gamma_1 \cdot \Big( Revals^-_{it-4} \times Large \, FA \,\, Share_{it-4} \Big) \\ \nonumber
    & + \gamma_2 \cdot \Big( Revals^-_{it-4} \times Large \, FL \,\, Share_{it-4} \Big) \\ \nonumber & + \mathbf{X}_{it-4} \mathbf{\Psi'} + \theta \cdot \Delta_{4q} \ln GDP_{t-4} + \alpha_i + \varepsilon_{it}.
\end{align}
where $Risk_{it}$ is either $Z\text{-}score_{it} = \dfrac{Cap_{it} + ROA_{it}}{\sigma_{4q}(ROA_{it})}$, any of its three components, or $\ln NPL_{it}$. In turn, $Cap_{it}$ is the capital to total assets ratio; $ROA_{it}$ is bank profitability, which we compute as the four-quarter sum of gross profit flows over total assets; $\sigma_{4q}(ROA_{it})$ is the four-quarter moving standard deviation of $ROA_{it}$ (for robustness, we also use eight- and twelve-quarter versions). $Large \, FA \,\, (FL) \,\, Share_{it}$ is a binary variable that equals 1 if a bank's foreign assets (foreign liabilities) to total assets ratio exceeds the sample mean and 0 otherwise. $\mathbf{X}_{it}$ includes other relevant bank-level characteristics (bank size, credit growth, liquidity) and the foreign assets and foreign liability shares themselves. $ \Delta_{4q} \ln GDP_{t}$ is the four-quarter log-difference of GDP (in constant prices). $\varepsilon_{it}$ is the regression error.

\begin{table}[h!] 
\footnotesize
\caption{Currency revaluations and bank stability}\label{tab:z_score}
\centering
\begin{tabular}{@{} p{0.4\textwidth} @{} 
>{\centering\arraybackslash}m{0.12\textwidth} @{} 
>{\centering\arraybackslash}m{0.12\textwidth} @{}
>{\centering\arraybackslash}m{0.12\textwidth} @{}
>{\centering\arraybackslash}m{0.12\textwidth} @{}
>{\centering\arraybackslash}m{0.12\textwidth} @{}
 }
\toprule[0.5mm]
	  & $Z\text{-}score_{it}$  & \multicolumn{3}{c}{Components of $Z\text{-}score_{it}$} & $\ln NPL_{it}$ \\    \noalign{\vskip 1mm}      \cmidrule(r){3-5}         
	&          & $\dfrac{EQ_{it}}{TA_{it}}$ & $ROA_{it}$ & $\sigma_{4q} (ROA_{it})$ & \\  \noalign{\vskip 1mm} 
\cmidrule(r){2-6} 
Explanatory variables ({\it lag = 4 quarters} )	& (1) & (2)       & (3)            & (4)         & (5)           \\ \noalign{\vskip 2mm} \hline \noalign{\vskip 2mm} 

$\beta$: Negative revaluations / Total costs, \% 
                    &    --15.710***&      --3.646** &      --1.227***&       0.314** &       0.415***\\
\hspace{2mm} ($Revals$)                     &     (5.829)   &     (1.818)   &     (0.316)   &     (0.124)   &     (0.133)   \\ \noalign{\vskip 4mm} 

$\gamma_1$: $Revals \, \times$ Large foreign asset share  &      13.102** &       3.041** &       0.530*  &      --0.314***&       0.009   \\
                    &     (5.745)   &     (1.526)   &     (0.317)   &     (0.116)   &     (0.129)   \\ \noalign{\vskip 2mm}

$\gamma_2$: $Revals \, \times$ Large foreign liability share  &       7.776   &       0.603   &       0.237   &      --0.065   &      --0.184   \\
                    &     (5.505)   &     (1.274)   &     (0.264)   &     (0.111)   &     (0.122)   \\ \noalign{\vskip 4mm}   
                    
Bank controls  & Yes & Yes & Yes & Yes & Yes \\ 
Bank FEs       & Yes & Yes & Yes & Yes & Yes \\ 
Macro controls  & Yes & Yes & Yes & Yes & Yes \\ \noalign{\vskip 2mm}

Obs        &    32,806  &    32,806 &  32,806   & 32,806 & 32,806 \\
Banks            &    1,080   &        1,080   &        1,080   &  1,080 & 1,080 \\
$R^2_{within}$   &       0.016   &       0.055   &       0.054   &       0.029   &       0.163    \\ \noalign{\vskip 4mm}

\it Test: $\beta$ + $\gamma_1$: & \it --2.608 & \it --0.605 & \it  --0.698*** &  \it 0.000 &  \it 0.423*** \\ 
        &  \it (5.846) &  \it (1.257) &  \it (0.267) & \it (0.109) &  \it (0.125) \\ \noalign{\vskip 2mm} 
        
\it Test: $\beta$ + $\gamma_1$ + $\gamma_2$: & \it 5.169 & \it --0.002 & \it  --0.460** &  \it --0.065 &  \it 0.240** \\ 
        &  \it (4.262) &  \it (0.926) &  \it (0.211) & \it (0.092) &  \it (0.100) \\        
        \noalign{\vskip 4mm}

\it 
Sample mean & \it 44.7 & \it 20.2 & \it 1.41 & \it 0.88 & \it 1.30 \\
\it Sample SD   & \it 43.6 & \it 14.2 & \it 2.10 & \it 0.87 & \it 0.95 \\
 
\noalign{\vskip 2mm}
\bottomrule[0.5mm] \noalign{\vskip 2mm}
\end{tabular}
    \begin{minipage}{1\linewidth} 
        {\it Note}: The table reports regressions of the Z-score of bank stability (1), its three components (2--4), and the log of bank NPLs (5) on the $Revals^-$ to total cost $TC$ ratio, controlling for bank size ($\ln TA$), the one-quarter log-difference of total (household and corporate) loans and its squared term ($\Delta_1 \ln LNS$ and $\Delta_1 \ln LNS^2$) to capture non-linearities in the relationship between credit expansion and bank stability, liquidity ratio ($LIQ/TA$, where $LIQ$ includes cash holdings and reserves), bank FEs, and the four-quarter log-difference of real GDP ($\Delta_4 \ln GDP$) to eliminate the effect of business cycle fluctuations. $Z\text{-}score_{it}$ is computed as the sum of bank capital to assets ratio $EQ_{it}/TA_{it}$ and bank four-quarter profitability $ROA_{it}$ divided by the four-quarter volatility of profits $\sigma_{4q} (ROA_{it})$. The larger the Z-score, the larger the distance to default. $ROA_{it}$ is the ratio of the annual gross profit to total assets, where the annual gross profit is the four-quarter moving sum of the quarterly flows of gross profits (i.e., before taxation and loan loss provisioning). $\sigma_{4q} (ROA_{it})$ is the four-quarter moving standard deviation of bank profitability. In the sample, the negative revaluations to the total costs variable has the mean $=0.29$ and (overall) standard deviation $=0.25$. {\it Large foreign asset (liability) share} is a binary variable that equals 1 if a bank's ratio of foreign asset holdings (foreign borrowings) to total assets is greater than the sample mean and 0 if else.  
        \\ ***, **, * indicate that a coefficient is significant at the 1\%, 5\%, 10\% level, respectively. Standard errors are clustered at the bank 
        level and appear in the brackets under the estimated coefficients.   \\
    \end{minipage}
\end{table}

The estimation results appear in Table \ref{tab:z_score}. In column (1), in which the dependent variable is the Z-score, we obtain a negative and highly significant coefficient estimate for $Revals^-$, a positive and significant estimate on its cross-product with the $Large \,\, FA \,\, share$ variable, and a positive but insignificant estimate on the $Large \,\, FL \,\, share$ variable. The first of the three estimates implies that the losses stemming from negative currency revaluations are especially damaging for domestic banks, i.e., those banks that have lower shares of either foreign assets or foreign borrowings (or both) on their balance sheets. Economically, a one standard deviation increase in  $Revals^-$ ($0.25$) is associated with a $3.9$ points decrease in the Z-score, which is equivalent to $9\%$ of the Z-score's standard deviation and is thus economically significant. Conversely, if a bank has a large share of foreign assets, it almost avoids the negative association between $Revals^-$ and Z-score, as our second estimate indicates. Indeed, in this case, the economic ``effect" of $Revals^-$ shrinks from $3.9$ to just $0.7$ ($=0.25 \cdot (-15.7 + 13.1)$) in terms of the Z-score and renders the impact economically insignificant. Foreign liabilities, though acting similarly, deliver no added value once foreign assets are controlled for.  

Regarding the three components of the Z-score, our further analysis shows that for domestic banks, $Revals^-$ are associated with significant contractions in bank capital to assets ratio (column (2)) and bank ROA (column (3)) and with significant spikes in the volatility of profits (column (4)). Again, as in the case of Z-score, we see that these negative relationships are offset, in part (3) or in full (2, 4), for banks possessing large shares of foreign asset holdings.

In column (5), where the dependent variable is the log of a bank's (domestic) NPLs, we obtain a negative and significant estimate of the coefficient on $Revals^-$ and insignificant estimates on its cross-products with both foreign operation variables. A one standard deviation rise in $Revals^-$ predicts a 0.1 log point spike in NPLs, which is equivalent to 11\% of one standard deviation of NPL  ($=0.415 \cdot 0.25 / 0.95$). An interpretation for this is that losses stemming from $Revals^-$ deteriorate profits and capital over assets, which, in turn, are likely to reduce a bank's ability to refinance their existing poorer-quality loans due to likely-binding regulatory constraints. 

Overall, we conclude that revaluations of foreign currency operations not only affect bank cost efficiency (levels and bank rankings) but they are also crucial to determining the efficiency of credit markets and stability of the banking system overall. The currency revaluations have a blurring effect: being accumulated in response to fluctuations of the nominal exchange rate, they contaminate the ``true" cost efficiency, create obstacles to expanding credit, and result in deteriorating bank stability through the reduction of both bank capital and ROA coupled with the increase in the volatility of ROA. Banks with larger foreign asset holdings are less sensitive to the negative effects of their $Revals$ at home compared to banks whose assets are more domestically oriented. 

\subsubsection{\it Does currency mismatch matter for high versus low capitalized banks? }
Finally, we address the concern that currency mismatches may have only a limited role, if any, for the financial stability of Russian banks, given that they hold on average relatively high capital buffers (see the discussion in Online Appendix \textcolor{red}{B}). High capital buffers, among other things, may well absorb any negative effects stemming from rising mismatches \citep{Abbassi2023}. In the previous section, we partly cover this issue by showing that larger foreign asset holdings attenuate the negative association between $Revals^-$ and financial stability. However, $Revals^-$ are just one side of the currency mismatch, and we are now going to investigate whether the currency mismatch matters for financial stability, especially in the case of higher (book) values of bank capital. 

\begin{figure}[h!]
    \centering
    \subfigure[Densities of bank capital in high and low states]{
    \includegraphics[width=0.47\textwidth]{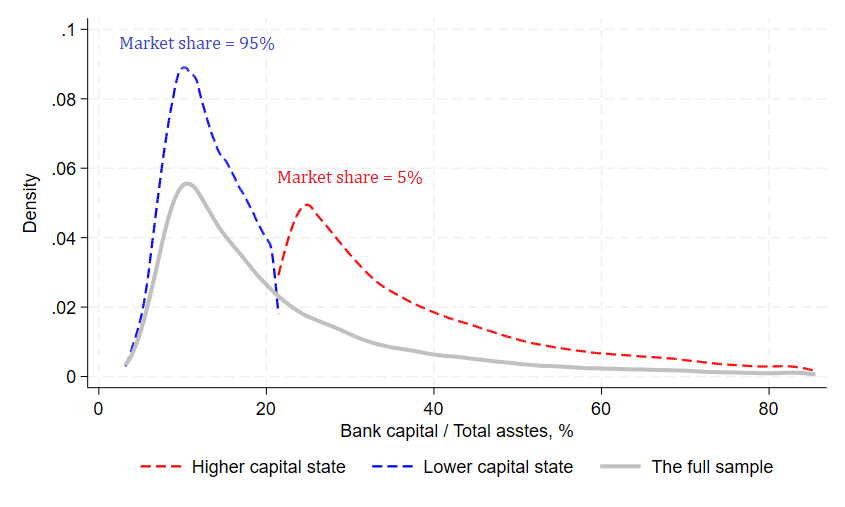}} 
    \subfigure[Net FX position and bank capital]{
    \includegraphics[width=0.47\textwidth]{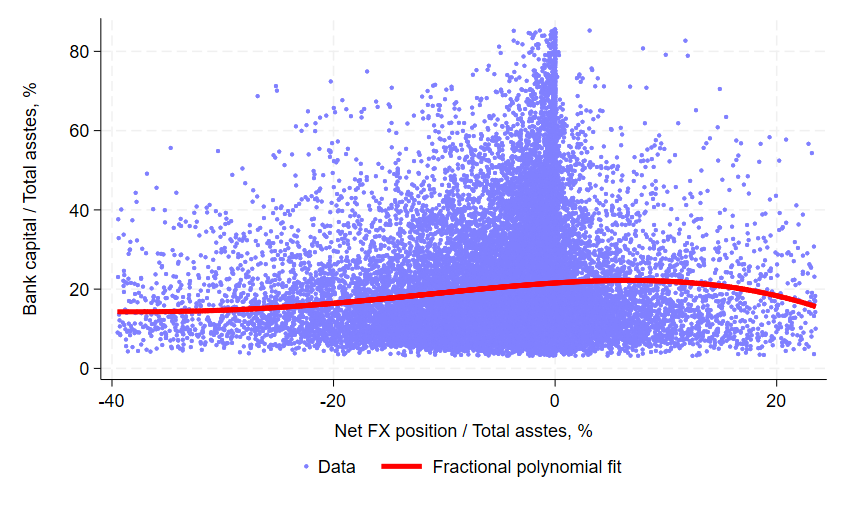}}
    
    \captionsetup{justification=centering,margin=2cm}
    \caption{Bank capital and net position in foreign currencies}
    \label{fig:CapNetFX}
\end{figure}

First, Figure \ref{fig:CapNetFX}({\it a}) documents that the overall distribution of banks' capital to assets ratio is left-skewed, ranging between 3 and 85\% and having an unconditional mean of 20\%. However, if we split the sample into two parts using this mean value, we observe that on average 539 banks with a market share of 95\% (in terms of total assets) are located in the {\it lower capital state}. Further, Figure \ref{fig:CapNetFX}({\it b}) indicates that the relationship between bank capital and net FX position is weak, that is the capital-to-assets ratios of banks that possess largely negative or largely positive currency mismatches are not substantially different from the capital-to-assets ratios of well-FX matched banks.

Second, we slightly modify Equation (\ref{eq:z_score}) from the previous section by replacing $Revals^-$ with the currency mismatch variable $Net\,FX$ (assets in foreign currencies net of liabilities in foreign currencies, as a \% of total assets) and introduce the interaction of $Net\,FX$ and the {\it High capital state} variable (equal to one if capital-to-assets ratio exceeds 20\%):
\begin{align} \label{eq:z_score_HighCap}
    Risk_{it} =& \, \phi \cdot Net\,FX_{it-4} + \pi \cdot \Big( Net\,FX_{it-4} \times High\,Capital_{it-4} \Big) \\ \nonumber
    & + Bank \, Controls_{it-4} \mathbf{\Psi'} + \theta \cdot \Delta_{4q} \ln GDP_{t-4} + \varepsilon_{it}.
\end{align}
We are agnostic about the sign of $\phi$, i.e., the relationship between $Net\,FX$ and $Risk$ for low-capital banks. It may be that $\phi>0$ because engaging in greater mismatch likely increases a bank's exposure to FX risk. However, as we discussed above, in countries with high political risks like Russia, banks may be willing to rely more on income from abroad, and the results in Table \ref{tab:z_score} reveal that banks with high shares of foreign assets are both more profitable and less volatile. In this case, we may anticipate $\phi<0$. We further hypothesize that $\pi<0$ if $\phi>0$, that is, high capital levels should absorb the negative effects of currency mismatch. We are agnostic about the sign of $\pi$ if $\phi<0$.

\begin{table}[h!] 
\footnotesize
\caption{Does currency mismatch matter for highly capitalized banks?}\label{tab:z_score_HighCAP}
\centering
\begin{tabular}{@{} p{0.4\textwidth} @{} 
>{\centering\arraybackslash}m{0.12\textwidth} @{} 
>{\centering\arraybackslash}m{0.12\textwidth} @{}
>{\centering\arraybackslash}m{0.12\textwidth} @{}
>{\centering\arraybackslash}m{0.12\textwidth} @{}
>{\centering\arraybackslash}m{0.12\textwidth} @{}
 }
\toprule[0.5mm]
	  & $Z\text{-}score_{it}$  & \multicolumn{3}{c}{Components of $Z\text{-}score_{it}$} & $\ln NPL_{it}$ \\    \noalign{\vskip 1mm}      \cmidrule(r){3-5}         
	&          & $\dfrac{EQ_{it}}{TA_{it}}$ & $ROA_{it}$ & $\sigma_{4q} (ROA_{it})$ & \\  \noalign{\vskip 1mm} 
\cmidrule(r){2-6} 
Explanatory variables ({\it lag = 4 quarters} )	& (1) & (2)       & (3)            & (4)         & (5)           \\ \noalign{\vskip 2mm} \hline \noalign{\vskip 2mm} 

Net FX position ($NetFX$) 
                    &      24.524***&      --0.983   &       0.009   &      --0.461***&      --0.071   \\
                    &     (7.676)   &     (1.307)   &     (0.403)   &     (0.162)   &     (0.190)   \\ \noalign{\vskip 2mm} 

High capital state  &       1.781   &       6.486***&      --0.183** &       0.241***&       0.171***\\
                    &     (1.697)   &     (0.369)   &     (0.075)   &     (0.032)   &     (0.034)   \\ \noalign{\vskip 4mm}
                    
$NetFX \, \times$ High capital state &       4.870   &       6.811*  &       0.635   &      --0.114   &      --0.462   \\
                    &    (16.861)   &     (3.587)   &     (0.824)   &     (0.328)   &     (0.382)   \\ \noalign{\vskip 4mm}
                    
Obs       &       25,637   &       25,637   &       25,637   &       25,637   &       25,633   \\
Banks           &         975   &         975   &         975   &         975   &         974   \\
$R^2_{within}$   &       0.019   &       0.130   &       0.032   &       0.036   &       0.120   \\ \noalign{\vskip 1mm}

\bottomrule[0.5mm] \noalign{\vskip 2mm}
\end{tabular}
    \begin{minipage}{1\linewidth} 
        {\it Note}: The table reports regressions of the Z-score of bank stability (1), its three components (2--4), and the log of bank NPLs (5) on currency mismatch proxied by the net position in foreign currencies ($NetFX$), controlling for the same characteristics as in Table \ref{tab:z_score} above. In the sample, the $NetFX$ variable has the mean $=-3.7$\% and (overall) standard deviation $=7.4$ pp. {\it High capital state} is a binary variable that equals 1 if a bank's ... and 0 if else.   
        \\ ***, **, * indicate that a coefficient is significant at the 1\%, 5\%, 10\% level, respectively. Standard errors are clustered at the bank level and appear in the brackets under the estimated coefficients.   \\
    \end{minipage}
\end{table}

The estimation results appear in Table \ref{tab:z_score_HighCAP}, which has the same structure as Table \ref{tab:z_score} in the previous section. First, in column (1), we obtain a positive and highly significant estimate of the coefficient on the $Net\,FX_{i,t-4}$ variable and a positive but small and insignificant estimate of the coefficient on the interaction of $Net\,FX_{i,t-4}$ and $High\,Capital_{i,t-4}$ variable. The first estimate implies that a hypothetical decrease in a bank's net FX position by one standard deviation (0.074) predicts a decline of the bank's Z-score by 1.8 points ($=24.524 \cdot 0.074$; equivalent to 4.1\% of the Z-score's standard deviation). The second estimate, in turn, indicates that higher capital levels are barely a remedy against this stability reduction: they do not absorb but rather amplify it. This ``amplification" is, however, very small in magnitude, i.e., only 0.4 points, or just 0.8\% of the Z-score's standard deviation, and is statistically insignificant. 

Very similar patterns arise from the other four columns as cross-products of high capital levels and the net FX variable appear insignificant (or at best marginally significant).\footnote{The results do not change qualitatively if we additionally control for the cross-products of $Net\,FX$ with either the indicator variable of high foreign assets share or the indicator variable of high foreign liabilities share (or both) as we did it in Table \ref{tab:z_score} in the previous section.} Of course, we do not claim causality here, and our predictive analysis just says that high-capital banks are not different from low-capital banks in terms of financial stability if both bank types have similar net FX positions.

\section{Conclusion} \label{sec:Conclusion}
This paper delivers systemic evidence in support of accounting for revaluations of FX assets and liabilities in the assessment of bank performance and credit market structure. When banks engage in FX mismatch, these revaluations tend to grow, especially during periods of currency crises. Using unique bank-level data from Russia where banks report these revaluations on both sides of P\&L accounts and the local currency is largely unstable, we find that FX revaluations substantially blur the actual landscape of the Russian banking sector in terms of bank cost efficiency. Removal of these revaluations dramatically alters the cost efficiency estimates and market structure evaluations. Our results are important beyond Russia and are likely to influence future studies on banking efficiency in other countries (e.g., EMEs) that are characterized by large holdings of FX assets and liabilities and encounter sizable fluctuations of the nominal exchange rate. 




\renewcommand{\thesection}{\Alph{section}} 
\renewcommand{\thesubsection}{\arabic{subsection}}
\renewcommand{\thesubsubsection}{\alph{subsubsection}}

\renewcommand{\thetable}{\Alph{section}.\Roman{table}}
\renewcommand{\thefigure}{\Alph{section}.\Roman{figure}}

\appendix
\renewcommand{\thesection}{Appendix \Alph{section}}
\renewcommand{\thesubsection}{\Alph{section}.\arabic{subsection}}

\clearpage
\section{Descriptive statistics} \label{app:descstats}
\setcounter{table}{0}

\begin{table}[h!] 
\footnotesize
\caption{Descriptive statistics for variables in cost frontier and in mean inefficiency regressions, 2005 Q1 -- 2020 Q2 ($N=38,484$)}\label{tab:desc_stats}
\centering
\begin{tabular}{lHcccc}
\toprule[0.5mm]
	                   & Obs & Mean & SD & Min & Max \\  \noalign{\vskip 1mm} 
	                   \cmidrule(r){2-6} 
	                   & (1) & (1)       & (2)            & (3)   & (4)                  \\ \noalign{\vskip 2mm} \hline \noalign{\vskip 2mm} 
\noalign{\vskip 2mm}

\multicolumn{6}{l}{\hspace{-2mm} {\it Inputs and outputs variables} } \\ \noalign{\vskip 2mm}
\hspace{4mm} Total loans (domestic and foreign, $Y_1$) &  38,656  &  48.8  &  561.6  &  0.00003  &  22,149.6 \\
\hspace{4mm} Total deposits (domestic and foreign, $Y_2$) &  38,656  &  38.7  &  480.6  &  0.00002  &  19,764.7 \\
\hspace{4mm} Fee and commission income ($Y_3$)  &  38,656  &  1.1  &  12.9  &  0.000001  &  664.3 \\ \noalign{\vskip 2mm}
\hspace{4mm} Wage rate ($w_1$) &  38,656  &  3.8  &  2.4  &  0.4  &  20.2 \\
\hspace{4mm} Capital expenses rate ($w_2$) &  38,656  &  23.0  &  15.9  &  2.4  &  83.2 \\ \noalign{\vskip 4mm}
\multicolumn{6}{l}{\hspace{-2mm} {\it Risk variables} } \\ \noalign{\vskip 2mm}
\hspace{4mm} Cash and reserves / total assets & 38,656 & 15.2 & 14.4 & 0.0008 & 98.7 \\
\hspace{4mm} Long-term loans to firms / total assets & 38,656 & 4.0 & 6.2 & 0 & 60.8 \\
\hspace{4mm} Long-term loans to households / total assets & 38,656 & 6.3 & 9.7 & 0 & 90.1 \\
\hspace{4mm} Equity capital / total assets & 38,656 & 20.0 & 14.0 & --42.2 & 98.1 \\
\hspace{4mm} Annual growth rate of total assets & 38,656 & 18.0 & 28.2 & --55.9 & 112.5 \\ \noalign{\vskip 4mm}
\multicolumn{6}{l}{\hspace{-2mm} {\it  Ownership Structure} } \\ \noalign{\vskip 2mm}
\hspace{4mm} Group 1: the Big-4 banks & 38,656 & 0.01 & 0.07 & 0 & 1 \\
\hspace{4mm} Group 2: other state-owned or -controlled banks & 38,656 & 0.01 & 0.12 & 0 & 1 \\
\hspace{4mm} Group 3: foreign-subsidiary banks & 38,656 & 0.05 & 0.22 & 0 & 1
\\ \noalign{\vskip 4mm}
\multicolumn{6}{l}{\hspace{-2mm} {\it  Regulatory Variables} } \\ \noalign{\vskip 2mm}
\hspace{4mm} Bank is under a bail-out & 38,656 & 0.01 & 0.11 & 0 & 1 \\
\hspace{4mm} Tight prudential regulation regime & 38,656 & 0.39 & 0.49 & 0 & 1 \\
 \noalign{\vskip 2mm}
\bottomrule[0.5mm] 

\end{tabular}
\end{table}

\end{document}